\documentclass[bibyear]{aa}  

\usepackage{natbib}
\usepackage{graphicx}
\usepackage[dvipsnames]{xcolor}
\usepackage{soul} 
\usepackage{txfonts}
\usepackage[colorlinks=true,
    linkcolor=blue,
    citecolor=blue,
    filecolor=magenta,      
    urlcolor=cyan]{hyperref}
\newcommand{\msun}{ M$_\odot$ }

\newcommand{\lsim}{\mathrel{\rlap{\lower 3pt \hbox{$\sim$}} \raise 2.0pt \hbox{$<$}}}
\newcommand{\gsim}{\mathrel{\rlap{\lower 3pt \hbox{$\sim$}} \raise 2.0pt \hbox{$>$}}}

\usepackage[skip=0.333\baselineskip]{caption}
\usepackage{subcaption}

\usepackage{booktabs}
\usepackage{ulem}
\usepackage{comment}
\usepackage{amssymb}
\begin{document} 

   \title{Can a time evolving, asymmetric broad line region mimic a massive black hole binary?}
   \authorrunning{E. Sottocorno et al.}
    \author{Erika Sottocorno
          \inst{1} \fnmsep\thanks{e.sottocorno2@campus.unimib.it}
          \and
          Mary Ogborn\inst{5}
          \and
          Lorenzo Bertassi\inst{1}
          \and 
          Fabio Rigamonti\inst{2,3,4}
          \and
          Matteo Bonetti\inst{1,2,3}
          \and
          Michael Eracleous \inst{5}
          \and
          Massimo Dotti\inst{1,2,3}
          }
          
   \institute{Università degli Studi di Milano-Bicocca, Piazza della Scienza 3, 20126 Milano, Italy
    \and
    INAF - Osservatorio Astronomico di Brera, via Brera 20, I-20121 Milano, Italy
    \and
    INFN, Sezione di Milano-Bicocca, Piazza della Scienza 3, I-20126 Milano, Italy
    \and
    Como Lake centre for AstroPhysics (CLAP), DiSAT, Università dell’Insubria, via Valleggio 11, 22100 Como, Italy
    \and
    Department of Astronomy and Astrophysics and Institute for Gravitation and the Cosmos, Penn State University, 525 Davey Lab, 251 Pollock Road, University Park, PA 16802, USA
    }

   \date{Received XXX; accepted YYY}

 
\abstract{Gas within the influence sphere of accreting massive black holes is
responsible for the emission of the broad lines observed in
optical-UV spectra of unobscured active galactic nuclei. Since the region contributing the most to the broad emission lines (i.e. the broad line region)
depends on the active galactic nucleus luminosity, the study of broad line reverberation to a varying continuum can map the morphology and kinematics of gas at sub-pc
scales. In this study, we modify a preexisting model for disc-like broad line regions,
including non-axisymmetric structures, by adopting an emissivity profile that mimics the observed luminosity-radius relation. This makes our implementation particularly well suited for the analysis of multi-epoch spectroscopic campaigns. After validating the model, we use it to check if strongly
non-axisymmetric, single broad line regions could mimic the short time-scale evolution expected from massive black hole binaries. We explore different orientations and anisotropy degrees of the broad line region, as well as different light curve patterns of the continuum to which the broad line region responds. Our analysis confirms that recently proposed algorithms designed to search for massive black hole binaries in large multi-epoch spectroscopic data are not contaminated by false positives ascribed to anisotropic broad line regions around single MBHs. 
}

   \keywords{techniques: spectroscopic -- galaxies: active -- galaxies: interactions -- quasars: emission lines -- quasars: supermassive black holes
}

   \maketitle
\section{Introduction}
Massive black hole (MBH) binaries (MBHBs) are expected to form during galaxy mergers \citep{1980Natur.287..307B}, and are expected to be the loudest sources of gravitational waves detectable in the nHz-mHz frequency range \citep[see][]{lisa1,lisa2,pta,2023A&A...678A..50E,2023ApJ...951L...8A,2023ApJ...951L...6R,2023RAA....23g5024X}.

Due to the small separations at which a MBHB forms \citep[$\approx 0.5$ pc for a binary with total mass $M_{\rm MBHB}=10^6$ \msun, see e.g.][]{2012AdAst2012E...3D}, resolving a binary (under the assumption that both MBHs are active) is extremely challenging. Effectively, this prevents any instrument from directly inferring through imaging the binary nature of an active galactic nucleus (AGN). Indeed, only one candidate has been identified through resolved radio imaging \citep[see][]{Rodriguez09}, justifying the active search for alternative ways to unveil MBHBs signatures \citep[see e.g.][for a comprehensive overview of the possible methods attempted so far]{Popovic2012,2019NewAR..8601525D, Bogdanovic2022}.
Still, MBHB signatures have been quite elusive because of the lack of unambiguous smoking guns of their effect on the EM radiation produced by an AGN. Indeed, the observational features used to identify MBHB candidates, including periodic continuum variability and shifted-asymmetric broad emission lines (BELs), can also be generated by AGN powered by single MBHs \citep[e.g.][]{ Vaughan2014, Liu2016,2022MNRAS.509..212D}. In many cases, long-term observations (performed over a period longer than the alleged orbital period of the binary candidate) can rule out false positives (i.e. single MBHs mimicking binary candidates). Unfortunately, the lengths of the observations required do not allow for such a test. For this reason \citet{Dotti} quantitatively study at length an alternative criterion, originally presented in a qualitative form in \cite{Gaskell88}, for the identification of sub-parsec binaries with orbital periods of $\gsim 10-100$ yr, for which long-term monitoring covering many orbital periods is challenging or unfeasible. This criterion is based on the uncorrelated variability of the broad line regions (BLRs), when they are still retained by each individual MBH in the binary. Due to the Doppler shift associated with their orbital motions, each BLR preferentially contributes to different velocities in the observed BEL profiles. A cross-correlation study of the blue and red sides of BELs therefore allows for the identification of (or validation of otherwise identified) MBHB candidates, since in the binary case no correlation is predicted, while for single MBHs high correlation is expected. 

In this paper, we check if the proposed method can give false positives (i.e. identify single MBHs as binaries) when a single BLR is sufficiently non-axisymmetric. In this situation, the BLR elements that contribute the most to the blue side of BELs (i.e. those moving toward the observer) can lie at different radii and react with different delays to variations of the continuum with respect to the gas contributing to the BEL red side. In order to verify the robustness of the observational test proposed in \cite{Dotti}, we model a single disc-like BLR featuring a non-axisymmetric perturbation based on the framework proposed by \cite{storchi03}. Building on that formalism and to maximize the realism of our model, we further modified the existing prescription to include the continuum luminosity dependence of the BLR radius \citep[see, e.g.][]{kaspi00,bentz09b}.
Specifically, our work will be divided as follows: in Sec.~\ref{sec:method} we construct the model to reproduce the flux associated with the specific system described above. In Sec.~\ref{sec:results}, we verify the physical consistency of our model with the observational results obtained from reverberation mapping campaigns, and we then perform a thorough parameter exploration to test the robustness of the observational test proposed in \cite{Dotti}. Finally, we draw our main conclusion in Sec.~\ref{sec:discussion}.

\section{Methodology}\label{sec:method}

\subsection{Geometry and Coordinate System}
\begin{figure}[h]
    \includegraphics[width = \linewidth]{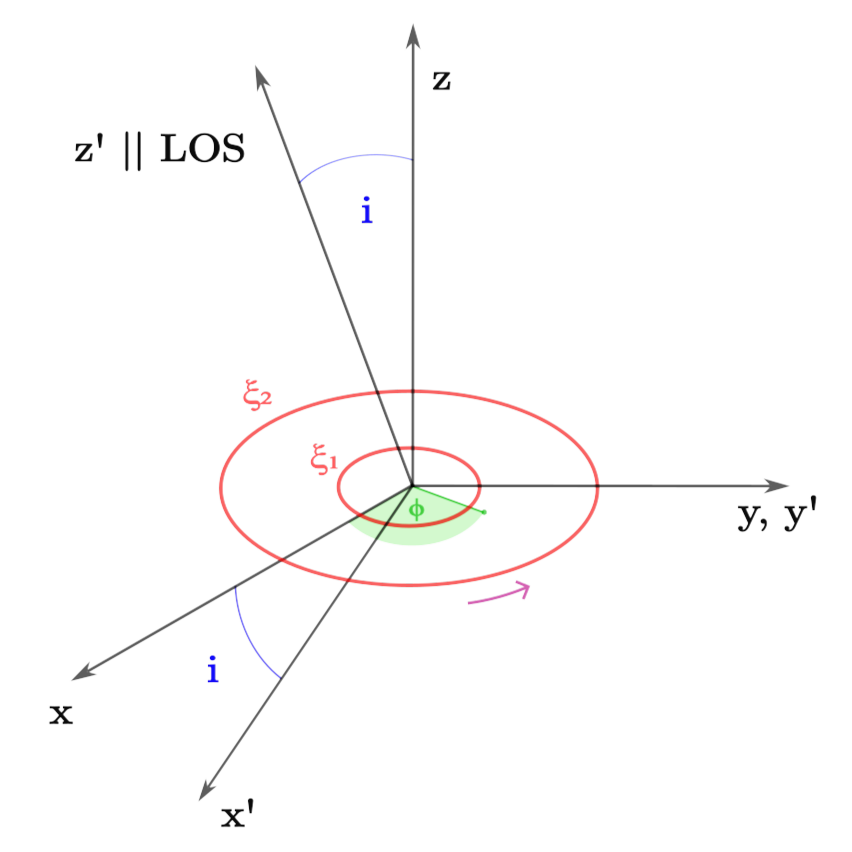}
    \caption{Geometry and coordinate systems used for the model \cite[see also][]{Chen_Halpern_Filippenko1989}. $\xi_1 $ and $\xi_2$ are respectively the inner and outer radius of the BLR, while $i$ is the inclination. The angle $\phi$ is measured with respect to the minor axis of the disc.}
    \label{fig:ref system}
\end{figure}

We assume a razor-thin, disk-like BLR geometry whose reference frame has been chosen as shown in Fig.~\ref{fig:ref system} \citep[see also][]{Chen_Halpern_Filippenko1989}. 
Here, the observer is chosen to be along the positive z-axis, while the disc is inclined with respect to the observer frame at an angle \textit{i}. The BLR disc major axis is aligned with the y-axis, while the angle $\phi$ is computed starting from the disc minor axis. The central object is assumed to be a black hole of mass $M = 10^{8}\, \rm{M_{\odot}}$\footnote{The results can be scaled for different MBH masses, by rescaling the properties of the BLR and the luminosity of the accreting MBH.} and it is located in the common origin of the reference systems. The disc is parameterized by an inner and an outer radius $\xi_1$ and $\xi_2$ where the dimensionless radius $\xi$, defined as $\xi = r/r_{G}$ with $r_G=GM/c^2$, has been used to facilitate analysis\footnote{Henceforth, to differentiate between the dimensionless radius and the actual radius, we respectively denote them as $\xi$ and $r$.}.

\subsection{Continuum Luminosity}\label{sec:luminosity}
The BLR elements respond to continuum variations with an average delay given by $\tau_{\rm{LTT}}$. The average delay can be obtained from the time lag between the fluctuations of the continuum and the integrated flux of an emission line, therefore it varies between emission lines. This delay is used to estimate the \textit{characteristic} size of the BLR $R_{\rm{BLR}}$, defined as $R_{\rm{BLR}} = c \, \tau_{\rm{LTT}}$
\citep[see][]{Blandford82, peterson93}.  To test the physical consistency of the proposed emissivity profile we study the BLR response to different continuum variations which we model by letting the intrinsic luminosity of the central MBH change with time. Specifically, to test different scenarios we studied the BLR reaction to an intrinsic luminosity parametrized by a step-function (see Eq.~\ref{eq:step_luminosity}):

 \begin{equation}
    \mathcal{L}_{\rm{step}}(t) =
        \begin{cases} 
          0 & t < \mathcal{T} \\
          1 & t \geq \mathcal{T},
       \end{cases}
       \label{eq:step_luminosity}
    \end{equation}

and a sinusoid (see Eq.~\ref{eq:sin_luminosity}):

\begin{equation}
    \mathcal{L}_{\rm{sin}}(t) = 10 + 5\sin{\left(\frac{2\pi}{\mathcal{T}}t\right)},
    \label{eq:sin_luminosity}
\end{equation}

where $\mathcal{T} = 40$ d for the step function and $\mathcal{T} = 120$ d for the sinusoidal continuum\footnote{This period was chosen in order for it to be in both cases smaller than the observational times considered in section \ref{sec:results} (set to $t = 150$ d).}, while the normalizations of the two light curves are completely arbitrary, as well as the total luminosity emitted by the BLR.

\subsection{BLR Emissivity}\label{par:emissivity}

As discussed above,
we test the robustness of the \cite{Dotti} test when applied to a single disc-like BLR
with strong non-axisymmetric perturbations. We assume a modified version of the formulation proposed by \cite{Gilbert1999} and \cite{storchi03}:

\begin{equation}
\begin{split}
 \epsilon (\xi, \phi; \xi_c, w) = &
\xi^{-1}\exp\left[-\frac{(\xi-\xi_c)^2}{2w^2}\right]\biggl\{1 +\\&\frac{A}{2}\exp\left[-\frac{4\log{2}}{\delta^2}(\phi - \psi_0)^2\right]+\\
 &+ \frac{A}{2}\exp\left[-\frac{4\log{2}}{\delta^2}(2\pi - \phi + \psi_0)^2\right]\biggr\},
\end{split}
\label{eq:emissivity}
\end{equation}
where the first term characterizes the axisymmetric part of the emissivity, while the other terms represent the deviation from axisymmetry caused by a perturbation with the form of a spiral arm. 


We modified the radial dependence of the emissivity in order to incorporate the observational relation between the BLR characteristic radius ($R_{\rm{BLR}}$) and the central source luminosity obtained by reverberation mapping campaigns \citep[see][]{kaspi00}. In particular, we will assume the scaling discussed in \cite{bentz09b}:
\begin{equation}
    R_{\rm{BLR}} \sim 11\,\rm{ld}\, \biggl(f_{\rm{Edd}}\frac{M}{M_\odot}\biggr)^{0.519},
    \label{eq:rad-lum relation}
\end{equation}
where $f_{\rm{Edd}}$ is the accretion bolometric luminosity normalized to the Eddington limit.
 The radial profile chosen in Eq.~\ref{eq:emissivity} ensures that the emissivity weighted average radius of the BLR $\xi_c$ obeys approximately the observed luminosity-radius relation in Eq.~\ref{eq:rad-lum relation} \footnote{Note that we approximated the exponent of the luminosity in Eq.~\ref{eq:rad-lum relation} to 0.5.}, as long as the centroid of the Gaussian term $\xi_c$ has the following dependence on the continuum luminosity:
\begin{equation}
    \xi_c = \xi_{c,0}\cdot\left(\frac{\mathcal{L}\left[t'(t_{\rm{obs}}, \xi, \phi)\right]}{\mathcal{L}_0}\right)^{0.5},
    \label{eq:r_c}
\end{equation}
where $\xi_{c,0}$ is the dimensionless average radius, $\mathcal{L}_0$ is the average luminosity, $t'$ is defined as

\begin{equation}
    t' = t_{\rm obs} - \xi \left[ \frac{r_{G}}{c} \cdot (1 + \sin{i} \cdot \cos{\phi})\right], 
    \label{eq:time delay}   
\end{equation}

to take into account the delayed response time for each BLR element, and $t_{\rm obs}$ is the time of observation. The width in the Gaussian term, $w$, is assumed to be equal to $w = \xi_c/2$.
\footnote{As commented in section~\ref{sec:luminosity}, for our test we set $\mathcal{L}_0$ in Eq.~\ref{eq:r_c} equal to 1 for the step function luminosity, and equal to 10 for the sinusoid luminosity.}

The angular dependent part of Eq.~\ref{eq:emissivity} is identical to the original \cite{storchi03}
proposal, where the second and the third exponentials are used to represent the decay of the emissivity of the spiral arm with azimuthal distance on either side of the ridge line, while the parameter \textit{A} depicts the brightness contrast between the spiral arm and the Gaussian emissivity underneath. The parameter $\delta$ is for the azimuthal full-width-at-half-maximum of the spiral arm, while the value $\phi - \psi_0$ denotes the azimuthal distance from the ridge of the spiral arm, defined by the following:
\begin{equation}
    \psi_0(\xi) = \phi_0 + \frac{\log_{10}{(\xi/\xi_{sp})}}{\tan{p}},
\end{equation}
with \textit{p} and $\xi_{sp}$ being, respectively, the pitch angle and the innermost radius of the spiral arm, while $\phi_0$ portrays the azimuthal orientation of the spiral pattern. In Appendix~\ref{app:emissivity}, we show how the BLR brightness changes with different combinations of the spiral arm parameters, while an application on real data is presented in \cite{Rigamonti2025}.

\subsection{Local Doppler Shift and Broadening}

After establishing the BLR emissivity, we take into account the effect of the local BLR bulk motion and turbulence on the Doppler shift and broadening of the emission line profiles. As the BLR rotates along the x-axis (see Fig.~\ref{fig:ref system}), the observer will perceive an increased or decreased frequency, depending on whether the emitting region is approaching or receding from them. Assuming the BLR elements to be on Keplerian circular orbits, the line-of-sight velocity is given by Eq.~\ref{eq:velocity}:

\begin{equation}
    v_{\rm{obs}} = -\sqrt{\frac{1}{\xi}}\sin{i}\sin{\phi}.
    \label{eq:velocity}
\end{equation}

Then, for the specific intensity of the line, we adopt the prescription proposed by \cite{Chen_Halpern_Filippenko1989} and \cite{Chen_Halpern1989}:

\begin{equation}
\begin{split}
    I_{\lambda} =\epsilon(\xi,\phi; \xi_c, w)&\cdot\frac{\mathcal{L}\left[t'(t_{\rm{obs}}, \xi, \phi)\right]}{\mathcal{L}_0}\cdot\\
    &\cdot\frac{1}{\sqrt{2\pi}\sigma_\lambda}\exp\left[-\frac{(\lambda - \lambda_e\cdot(1+v_{\rm{obs}}/c))^2}{2\sigma_\lambda^2}\right],
\end{split}
    \label{eq:intensityde}
\end{equation}

where $\epsilon(\xi, \phi; \xi_c, w)$ is the emissivity defined by Eq.~\ref{eq:emissivity}, $\mathcal{L}\left[t'(t_{\rm{obs}}, \xi, \phi)\right]$/$\mathcal{L}_0$
represents the apparent luminosity pattern of the disc by the continuum source, which sets the local effective emissivity.
A local broadening is included as a Gaussian profile with a broadening parameter $\sigma_\lambda$ set to $1200$ km/s in the model, as in the original \cite{storchi03} implementation.
In the following analysis, we will model the response of the H$\alpha$ line, but the same test can be performed for different BELs as long as the parameters of the model are properly tailored. Therefore, the wavelength of the line in the emitter frame is set to $\lambda_e = 6563$ \AA.
Finally, we stress that the line intensity profile in Eq.~\ref{eq:intensityde} has an explicit dependence on the impinging continuum luminosity, to allow the BEL to increase or decrease its luminosity in response to the continuum modulations.

We note that not all the relativistic processes considered in the previous implementations \citep[e.g.][]{Chen_Halpern1989, Chen_Halpern_Filippenko1989} have been taken into account in the current model. In particular, the effect of gravitational redshift, light bending, and any effect associated to radiative transfer processes within the BLR are not taken into consideration at the current stage. A brief comparison with a similar implementation including such effects is presented in Appendix~\ref{app:model_comparison}. Although such effects have a small influence on the line profiles, they do not influence the results of the test we perform in this work.

\subsection{Binary Test through Cross-Correlation Analysis}\label{sec:fuvt}

From an observational point of view, the measurement of the BLR physical size utilizes a collection of several spectra taken at different times containing at least one BEL. The time delay between the AGN continuum and the reprocessed light from the BLR is typically estimated by comparing their respective light curves. Multiple approaches exist to estimate such time delays efficiently \citep{peterson04,pancoast14,Zu2011,Raimundo2020,Donnan2021,Nunez2023}. Most of them, such as the method of \cite{peterson04} employed here, are based on the maximization of the cross-correlation between the two light curves.

\cite{Dotti} proposed a test which, exploiting RM observations, identifies MBHB at separations large enough for the two MBHs to still retain their individual BLR (i.e. a separation larger than the Roche lobe of the binary system). In our methodology (dubbed fast uncorrelated variability test - FUVT) the variable part of a broad emission line is divided into eight equal flux wavelength bins identifying seven $\bar{\lambda}$.
For each of the $\bar{\lambda}$ we divide the spectrum of the selected BEL into a red and blue component (for wavelengths longer and shorter than $\bar{\lambda}$, respectively). We then compute the light curves of those components, perform the cross-correlation between the two, and measure the maximum of the cross-correlation for each $\bar{\lambda}$ \citep[see][for further details]{dotti23}. At this point, if the cross-correlation is low for all the $\bar{\lambda}$ (especially for those concentrated near the bulk of the line where the S/N is not likely to affect the measurement), there might be an indication of the presence of a MBHB. Indeed, in the case of two detached BLRs orbiting around the common center of mass of the system, it is expected for each one of them to predominantly contribute to the red or blue light curve. Moreover, since the two BLRs are reprocessing the light coming from two different ionizing sources, their red and blue light curves will show much less correlation compared to the case of a single MBH.

\section{Results}\label{sec:results}

In this section, we show how the line profile at each time and its overall time evolution depends on the parameters assumed for the BLR emissivity. Subsequently, we will focus on the cross-correlation analysis presented above, performed using the \textsc{Python} adaptation \citep[see][]{Sun} of the code originally developed by \cite{peterson04}. We will first check the ability of our implementation to reproduce the luminosity-radius relation, by computing the cross-correlation between the total flux of the BEL and the continuum. Then, we compute the cross-correlation between the red part and the blue part of the line for a statistically relevant number of BLR profiles, to test whether the method could wrongly associate to these scenarios the presence of a MBHB in the context of the FUVT described above.

\subsection{Emission Line Profiles}\label{Sec:fluxprofile}

\begin{figure*}[h]
    \centering
    \includegraphics[width = \linewidth]{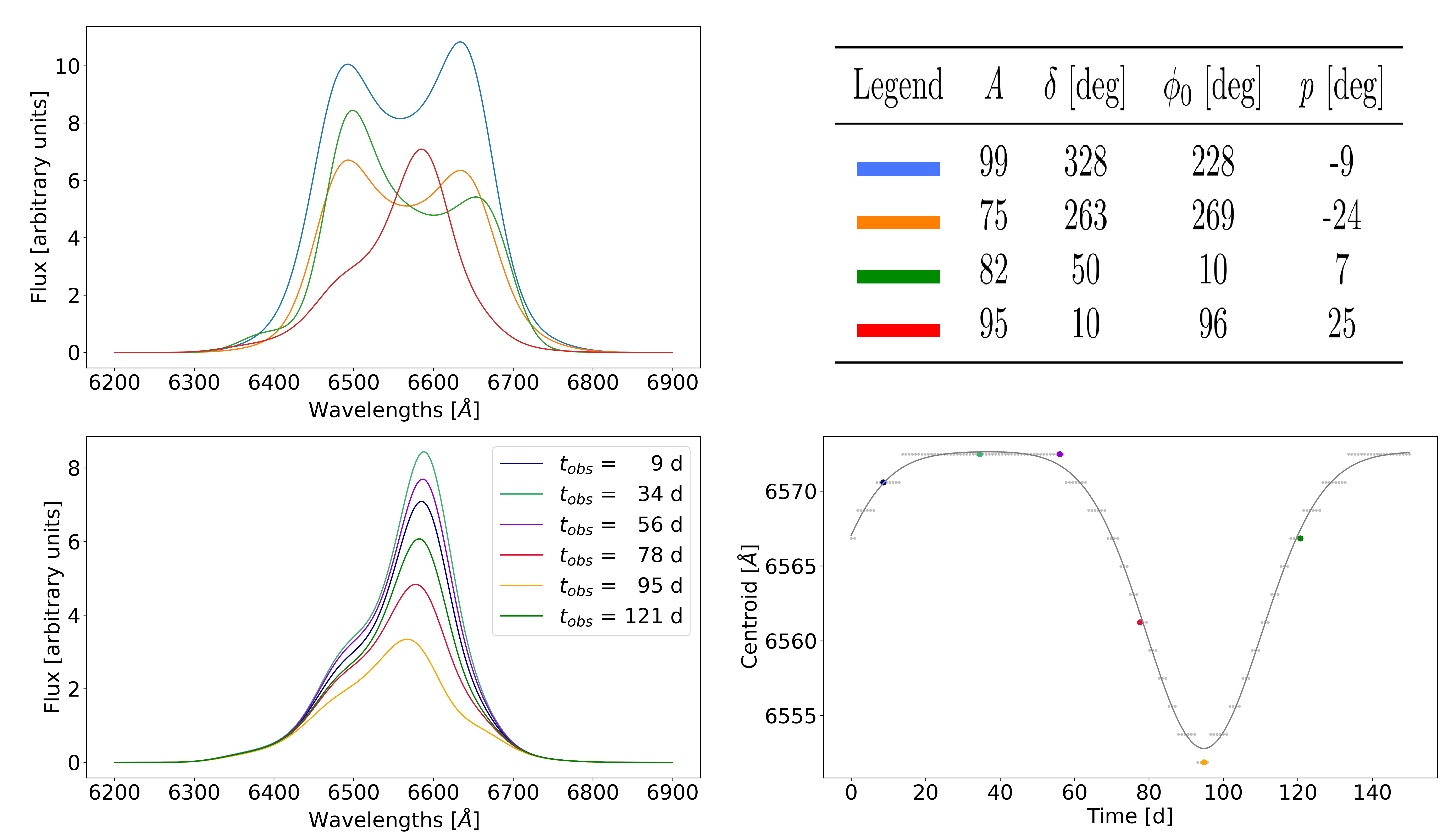}
    \caption{Example of line profiles. On the first row, different BELs obtained at the same time of observation for different sets of parameters (listed on the upper right legend). On the second row, the evolution of the BEL obtained by the third set of parameters (i.e. the ones associated to the green emission profile in the first row) is shown at different observation times (left) with the corresponding centroid (right).}
    \label{fig:flussi emissività gaussiana}   
\end{figure*}

The upper left panel of Fig.~\ref{fig:flussi emissività gaussiana} shows the H$\alpha$ BEL for different sets of emissivity parameters (listed on the top right corner of the figure) under the assumption of a sinusoidal variation of the continuum (see Eq.~\ref{eq:sin_luminosity}). Each emission line represents the same snapshot (i.e. the flux is taken at the same time of observation). As expected, the spiral perturbation introduces asymmetries in the emission line profile based on its different properties. As seen in Fig.~\ref{fig:flussi emissività gaussiana}, we allow the parameter $A$ to vary up to 100. This approach enables us to evaluate the reliability of the FUVT by considering also strongly non-axisymmetric BLR. However, we stress that these values are higher than the ones used to fit observed line profiles. For instance, in \cite{Storchi17} and \cite{Ward24} reasonable fits to observed line profiles were obtained constraining $A$ to values below 10. Hence, we extend the variation of $A$ up to 100 to test the FUVT reliability under extreme conditions.

\begin{figure*}[h]
    \includegraphics[width = \linewidth]{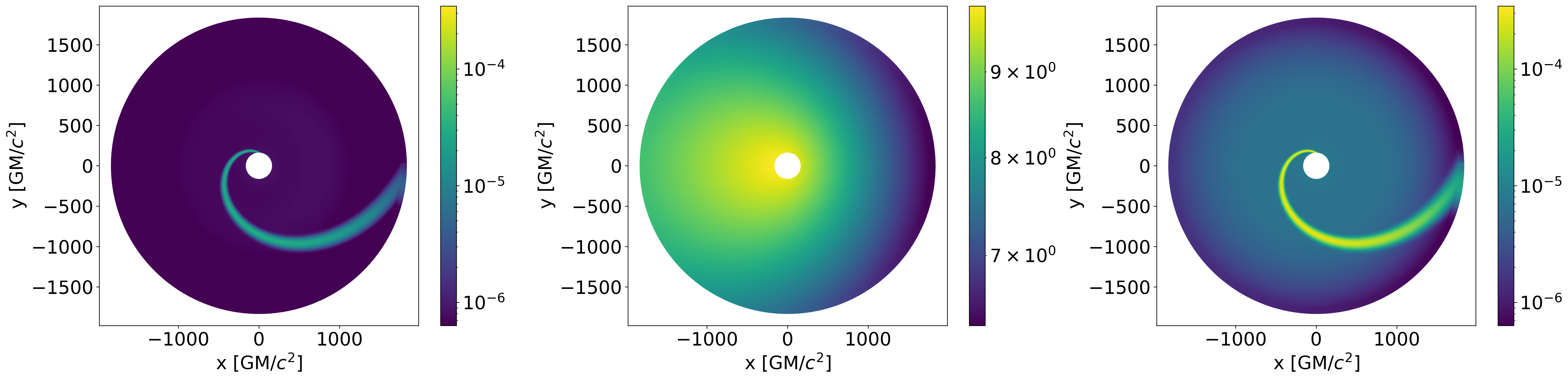}
    \caption{Illustration of how the disk responds to the variable ionizing continuum for the set of parameters $A$ = 89, $\delta$ = 10°, $\phi_0$ = 96, $p$ = 25°.  The line of sight is assumed to be inclined by  $i$ = 30° with respect to the normal to the disk in the direction of the negative $x$-axes. From left to right: the intrinsic emissivity map, the instantaneous illumination map and the instantaneous brightness map. As it can be noticed, the impact of $\mathcal{L}(t')$ leads at the assumed time ($t=0$) to a greater enhancement of the BLR emissivity on the} near side (i.e. closer to the observer) compared to the far side.
    \label{fig:trittico_BLR}   
\end{figure*}

In the second row, we fix the BLR parameters at those identified by the red colour in the legend and show the evolution of the BEL as a function of time (i.e. from $t_{\mathrm{obs}}=0$ to $t_{\mathrm{obs}}= 150$ d). Specifically, the bottom left panel shows the line profile at four different snapshots reported in the panel legend. As seen in this panel, the height of the profile varies with the observation time. This happens because, in our model, the line profile is linearly dependent on the continuum, here assumed to be a sinusoidal continuum with a period of $\mathcal{T}$ = 120 d. Therefore, the profile follows a similar behaviour. On the bottom right panel, we report the evolution with time of the line centroid computed by considering the wavelength at which the line is split into two equal flux regions\footnote{Note that the centroid shifts up or down in wavelength as the line integrated flux increases or decreases, respectively.}. As noted in the introduction, the presence of an asymmetric BEL with line centroid (either the peak or flux weighted wavelength) shifted with respect to the host redshift has been used as a tool to identify MBHB candidates \citep[see, e.g.][for examples of systematic searches]{Tsalmantza11, Eracleous12}. The shift of the line centroid in time can be used to distinguish between false binary candidates and true MBHBs \citep[e.g.][]{Decarli13, Runnoe15, Runnoe17}. The example in Fig. 2 demonstrates that small shifts in the BEL centroid can be obtained even with a single MBH irradiating a single non-axisymmetric BLR. However, these shifts are of the order of a few hundred kms$^{-1}$ and they would tend to introduce noise ("jitter") in the long-term radial velocity curves that track the line peak, as discussed in \cite{Runnoe17}. In the following we will demonstrate that for large separation binaries, as the one considered in the current study, the test proposed in \cite{dotti23} can distinguish between single and binary MBHs, while at smaller separations ad-hoc tests have to be devised to prevent the misidentification of single MBHs as binaries, as discussed in Bertassi et al.2025a (submitted). Finally, in Fig.~\ref{fig:trittico_BLR} we show the intrinsic emissivity map $\epsilon(\xi, \phi; \xi_c, w)$ for this set of parameters, the instantaneous illumination pattern for the sinusoidal input light curve (i.e. $\mathcal{L}(t')$ where $t'$ is given by Eq.~\ref{eq:time delay}) and their corresponding product, which is the instantaneous brightness map (i.e. the input to the line profile computation) at a fixed observation time t = 0 d. This is done to illustrate how the system responds to the ionizing continuum.

\subsection{Reverberation Mapping: Continuum versus Total Integrated Flux of the BEL}

Here, we show how well our model is able to reproduce the luminosity-radius relation obtained through reverberation mapping campaigns. In order to test our model, we first fixed the spiral parameters to constant values reported in Tab.~\ref{tab:spiral parameters 1}, along with the fixed values of inclination, and inner and outer radii of the disc.

\begin{table}[h]
    \centering   
    \begin{minipage}[t]{0.2\textwidth}
        \centering
        \begin{tabular}{cc}
            \midrule
            Parameter & Value \\
            \midrule
            \textit{i} [deg] & 30 \\
            $\xi_{1}$ & 200 \\
            $\xi_{2}$ & 1800\\
            $\xi_{sp}$ & 200 \\
            \midrule
        \end{tabular}
    \end{minipage}
    \begin{minipage}[t]{0.2\textwidth}
        \centering
        \begin{tabular}{cc}
            \midrule
            Parameter & Value \\
            \midrule
            $\phi_0$ [deg] & 120 \\
            $\delta$ [deg] & 110 \\
            \textit{A} & 2 \\
            \textit{p} [deg] & -25 \\
            \midrule
        \end{tabular}
    \end{minipage}
    \caption{Set of the spiral disc model parameters used to check the physical consistency of our BLR model.}
    \label{tab:spiral parameters 1}
\end{table}

\begin{figure}[h]
    \includegraphics[width=\linewidth]{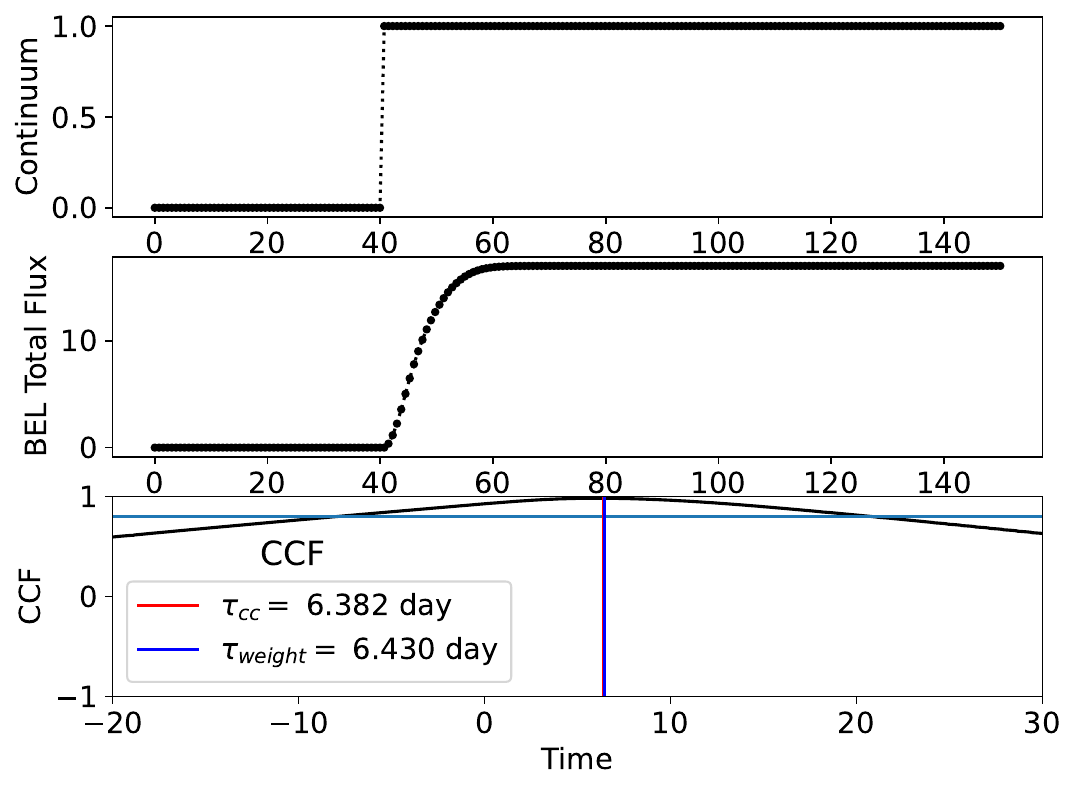}
     \caption{Cross-Correlation Analysis, assuming a step function continuum (i.e. Eq.~\ref{eq:step_luminosity}). On the upper panel, the continuum luminosity. On the middle panel, the total integrated fluxes of a BEL.
     Finally, on the lower panel, the cross-correlation computations.}
    \label{fig:riga vs continuo}
\end{figure}

\begin{figure}[h]
    \includegraphics[width=\linewidth]{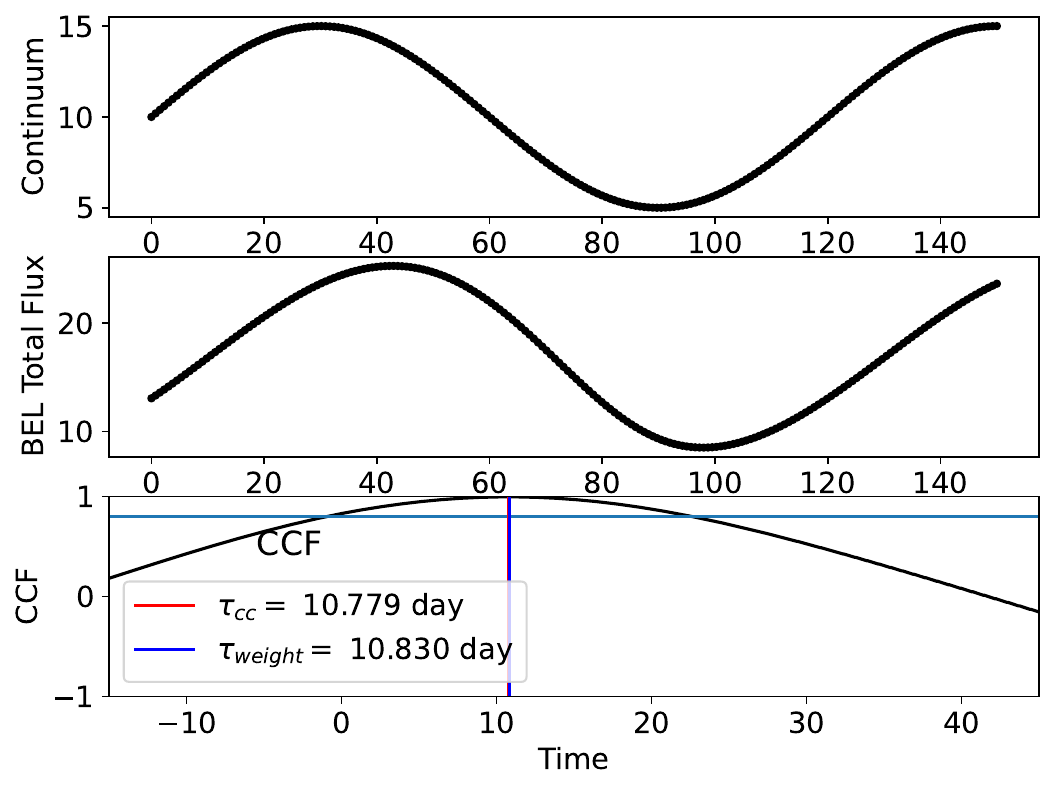}
     \caption{Same as Fig.~\ref{fig:riga vs continuo} assuming a sinusoidal continuum (i.e. Eq.~\ref{eq:sin_luminosity}).}
    \label{fig:riga vs continuo_sinusoid}
\end{figure}

Then, we let $\xi_{c,0}$ vary between $\xi_1$ and $\xi_2$\footnote{Recall that $\xi_1$ and $\xi_2$ are the inner and outer radii of the disk like BLR.}. For each value of $\xi_{c,0}$, we computed the line profile for 200 different observational times between 0 and $t = 150$ d. We then used them to construct the light curve for the total integrated line flux. 
Finally, we calculated the cross-correlation between the integrated line and the continuum considering both the step function and the sinusoidal continuum (Fig.~\ref{fig:riga vs continuo} and Fig.~\ref{fig:riga vs continuo_sinusoid}, respectively). In both figures we show from top to bottom: the continuum luminosity (either Eq.~\ref{eq:step_luminosity} or Eq.~\ref{eq:sin_luminosity}), the corresponding  BEL response, and the cross-correlation between the continuum and the integrated line as a function of time. In the bottom panel the blue line represents the time delay maximizing the cross-correlation ($\tau_{cc}$), while the red line  shows the averaged time delay ($\tau_{\rm weight}$) estimated using only values with a corresponding cross-correlation value exceeding a threshold equal to $0.8$. As can be observed by the legend reported in the panel the two methods give consistent results. The time delays are then used to compute $R_{\rm BLR}$. 

\begin{figure}[h]
    \centering
    \includegraphics[width=0.49\textwidth]{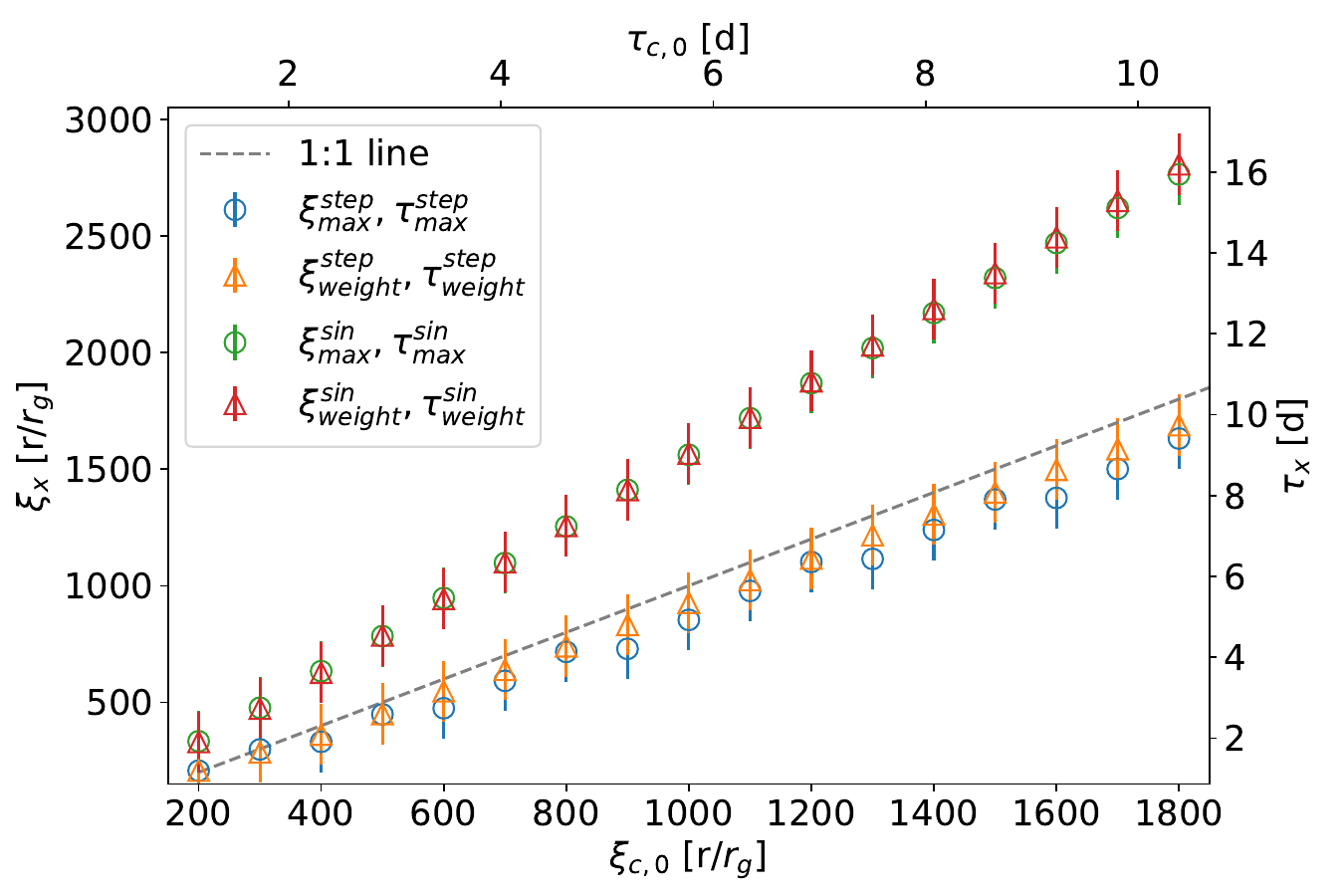}
    \caption{Comparison between the characteristic radius of the BLR obtained by applying reverberation mapping techniques and the radius obtained from the luminosity-radius relation. we show with blue and orange markers the results from the step-function continuum luminosity, while we use green and red markers for the sinusoidal continuum. In both cases, the radius has been estimated considering the maximum of the cross-correlation as well (represented by circular markers) as the weighted average of the time delays (represented by a triangular markers) (see Sec.~\ref{sec:results} for further details).}
    \label{fig:time delays}
\end{figure}

In Fig.~\ref{fig:time delays} we show the characteristic size of the BLR estimated from the cross-correlation analysis for both the step function and the sinusoidal continuum. The circular and triangular shapes refer to the specific approach considered for estimating the characteristic size of the BLR (maximum or weighted average, respectively), while the different colours represent different underlying continua. In particular, blue and orange markers are used for the step function continuum, while red and green markers are associated with the sinusoidal continuum. The errors are estimated based on the interpolation time step between the evaluated time delays. The additional vertical and horizontal axes report the time delays in units of days.

As it can be seen from Fig.~\ref{fig:time delays}, the values obtained for the sinusoidal continuum are always higher with respect to the step function ones. 
Our analysis shows that the exact value of the ratio between the characteristic size of the BLR obtained through the reverberation mapping and the emissivity averaged radius $\xi_{c,0}$ depends on the time-evolution of the continuum. This result is already known, and it is discussed from an analytical standpoint in \cite{krolik01}. Nonetheless, we find a linear relation between $R_{\rm BLR}$ and  $\xi_{c,0}$. Such a linear scaling maps into the expected dependence of $R_{\rm BLR}$ and the continuum luminosity, allowing for the use of the BLR model introduced for the analysis of multi-epoch spectra.

\subsection{Reliability of the FUVT in the asymmetric BLR scenario}\label{blue vs red}
Once we verified the physical consistency of our model, in order to investigate the robustness of the FUVT method, we computed the cross-correlation between the blue and red fluxes of the emission line for a large number of different BLR scenarios. Specifically, contrary to what was done in Sec.~\ref{Sec:fluxprofile}, we fixed $\xi_{c,0} = 1200$, considered the sinusoidal continuum, and let the spiral parameters vary, creating 10 000 different BLR configurations corresponding to 10 000 BEL profiles. We assumed the parameter ranges to be $\delta, \phi_0 \in [0, 2\pi]$, $p \in [-\pi/2, \pi/2]$, and $A \in [0, 100]$, respectively.\footnote{Specifically, the distributions for $\delta$, $\phi_0$, and $p$ are all uniform, while a log-uniform distribution is used for $A$.} As mentioned earlier, the broader range for $A$ with respect the one resulting from fits of observed line profile is chosen to account for strongly non-axisymmetric BLR, allowing us to test the FUVT reliability under these extreme scenarios as well.
For each BLR profile, we first computed the line profile for 175 different observational times between $t_{\mathrm{obs}} = 0$ and $t_{\mathrm{obs}} = 150$ d. Due to the high number of scenarios to be processed, we used an NVIDIA TESLA P100 GPU - provided by \hyperlink{Kaggle}{Kaggle}\footnote{\hypertarget{Kaggle}{\url{https://www.kaggle.com/}}} - to optimize the efficiency of our code by parallelizing the generation of fluxes. This approach allowed us to generate 175 line profiles (i.e. one scenario) in 6 seconds. On a standard CPU (i.e. AMD Ryzen 7 3700U), the same computation is performed in 7 minutes.

\begin{figure}[h]
    \centering
    \includegraphics[width = \linewidth]{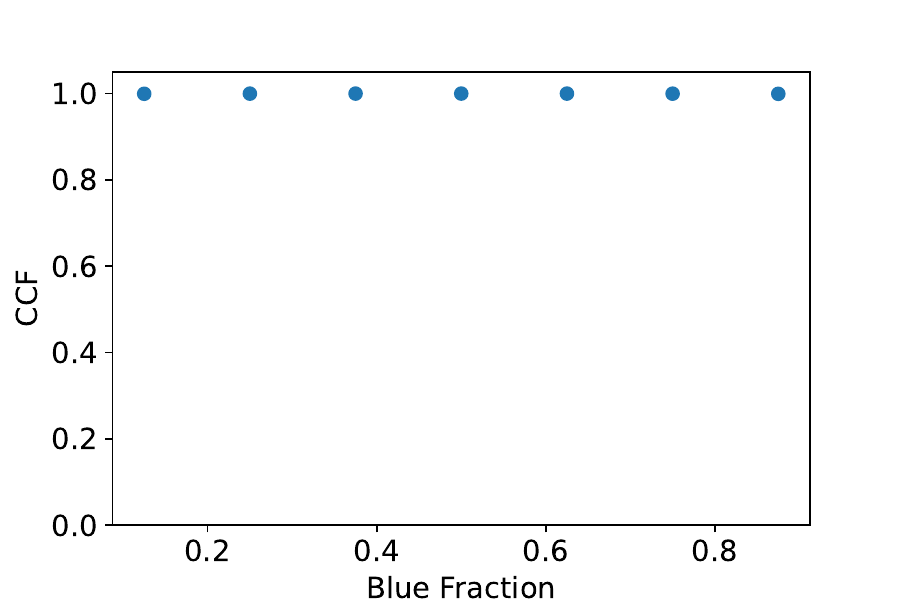}
    \caption{Example of CCF results for a random set of BLR parameters, as a function of $\bar{\lambda}$, corresponding to different fractions of the total BEL luminosity in the blue part of the line.}
    \label{fig:cross correlation}
\end{figure}

Then, following \cite{Dotti}, we determine the cross-correlation dependence on the wavelength at which
the division of the line is performed, as summarized in Sec.~\ref{sec:fuvt}. An example of the seven distinct cross-correlation values obtained for a random set of BLR parameters is shown in Fig.~\ref{fig:cross correlation}.

\begin{figure}[h]
    \centering
    \includegraphics[width = \linewidth]{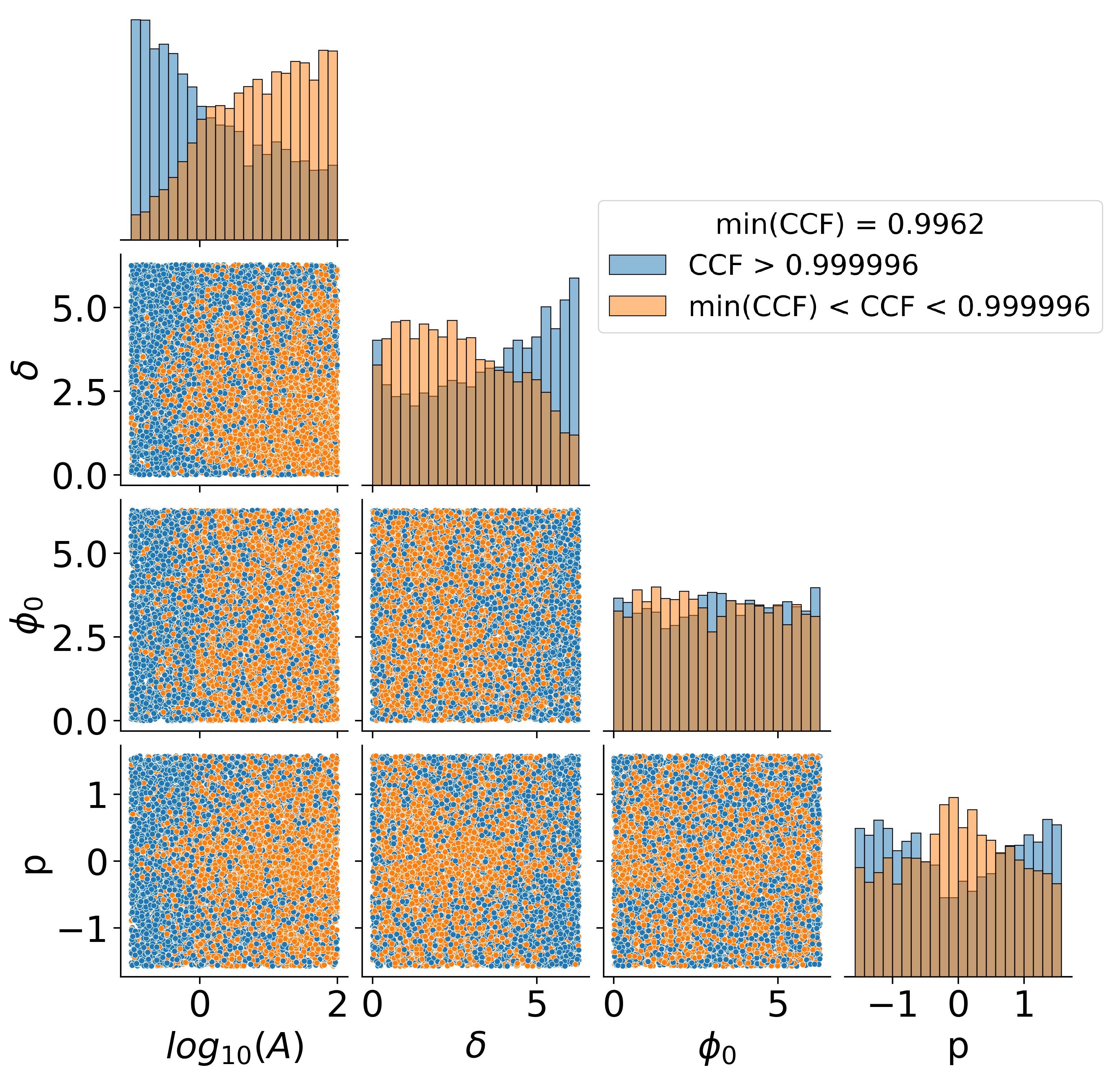}
    \caption{Pairplot showing the cross-correlation value distribution in the spiral parameters space. The obtained values are always higher than $\mathrm{min}(CCF) = 0.9944$. Thus, the cross-correlation consistently approaches a value close to unity. We categorized the data into two classes, using the median value as the cut-off.}
    \label{fig:pairplot}
\end{figure}

For each BLR profile, we considered the maximum value obtained for the cross-correlation, to analyze the predictions that would be obtained from the method. As it can be seen from the corner plot shown in Fig.~\ref{fig:pairplot}, all the different scenarios returned a maximum for the cross-correlation on the order of unity, with a minimum value given by $ \mathrm{min}(CCF) = 0.9944$.

As our intent is to verify the impact of the intrinsic properties of the BLR on the red-blue cross-correlation, we worked under the assumption of zero uncertainties on the BEL profiles. However, as discussed in \cite{Dotti}, observational errors on the data might reduce the red-blue cross-correlation, both for observations of single or binary systems. A detailed study of the effect of noise is the topic of a follow-up study (Bertassi et al. 2025b in prep.).
Here we only show an example of the same analysis discussed above, in which we assumed the BLR parameters in Tab.~\ref{tab:spiral parameters noise} and $\xi_{c,0} = 1200$, and we added to the resulting BEL profile different levels of Gaussian noise.

\begin{table}[h]
    \centering
    \begin{minipage}[t]{0.2\textwidth}
        \centering
        \begin{tabular}{cc}
            \midrule
            Parameter & Value \\
            \midrule
            \textit{i} [deg] & 30 \\
            $\xi_{1}$ & 200 \\
            $\xi_{2}$ & 1800\\
            $\xi_{sp}$ & 200 \\
            \midrule
        \end{tabular}
    \end{minipage}
    \begin{minipage}[t]{0.2\textwidth}
        \centering
        \begin{tabular}{cc}
            \midrule
            Parameter & Value \\
            \midrule
            $\phi_0$ [deg] & 120 \\
            $\delta$ [deg] & 16 \\
            \textit{A} & 100 \\
            \textit{p} [deg] & -5 \\
            \midrule
        \end{tabular}
    \end{minipage}
    \caption{Set of the spiral disc model parameters used to study the cross-correlation dependence on the noise. These remain constant across the various cases, with the noise amplified by adjusting the variance of the superimposed Gaussian fluctuations.} 
    \label{tab:spiral parameters noise}
\end{table}

\begin{figure*}[h]
    \centering
    \includegraphics[width=\linewidth]{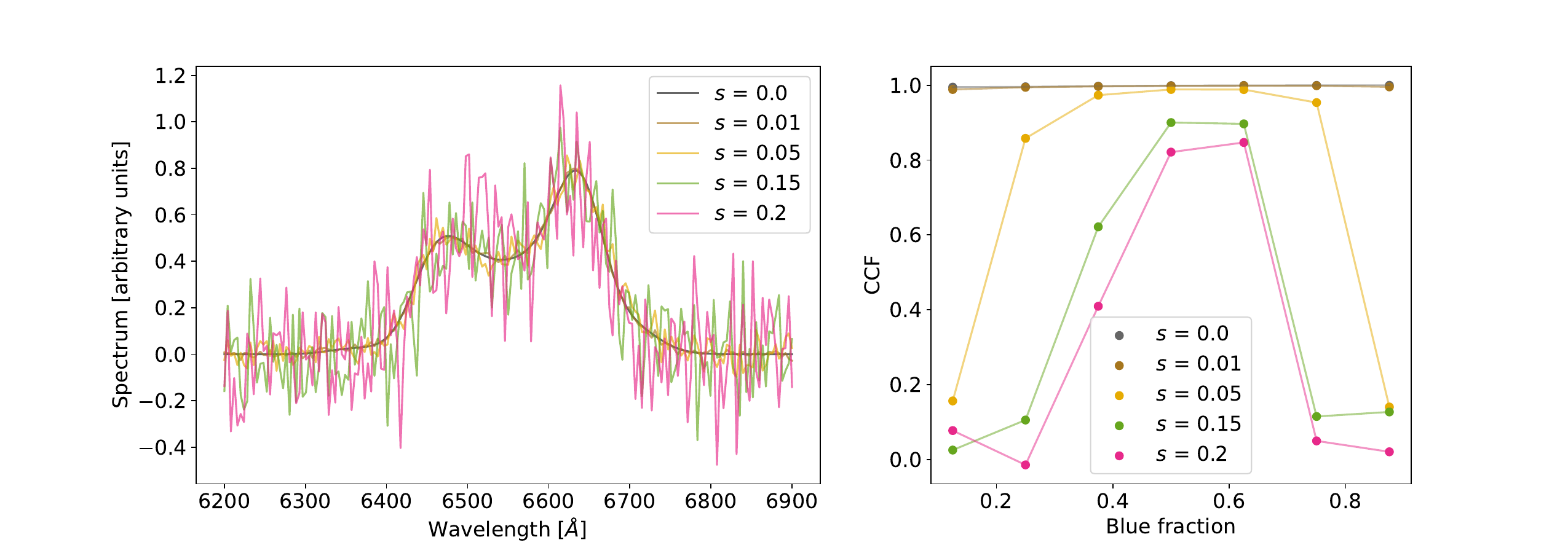}
    \caption{On the left panel, noise implementation on the profile obtained using the set of parameters listed in Tab.~\ref{tab:spiral parameters noise} at $t = 0$. Here, we assume Gaussian noise with different values for $s$, as indicated in the inset. We stress that only the BEL is plotted, while an AGN spectrum in the broad H$\alpha$ spectral region would feature, at the very least, the AGN and the host galaxy continua. The negative values of the flux are therefore not unphysical, and should be interpreted as uncertainties on the total (positive) flux. On the right panel, CCF results as a function of the noise. The value of the cross-correlation decreases as the amplitude of the noise increases. The maximum of the cross-correlation remains higher than $0.8$ for all the cases considered, in which $s$ remains smaller than the peak value of the BEL flux.}
    \label{fig:cc_fluxes_noise}
\end{figure*}

Specifically, at each time and for each wavelength we added to the flux predicted by the BLR model an independent sampling from a Gaussian centred at 0 and with different values of $s$, as illustrated in the left panel of Fig.~\ref{fig:cc_fluxes_noise} for $t = 0$.

The cross-correlation between the blue and the red part of the BEL is then computed as in the previous (noiseless) case. As shown in the right panel of Fig.~\ref{fig:cc_fluxes_noise}, the value of the cross-correlation is indeed modified by the addition of the noise, with the profile of $CCF$ as a function of the BEL cut getting similar to the one observed in most of the single MBH cases analyzed in \cite{Dotti}: the $CCF$ remains high ($\gsim 0.8$) when dividing the BEL in two close-to-equal-flux parts, while, it decreases significantly toward no correlation when one side of the line accounts for a small fraction of the overall flux and is, therefore, strongly affected by the noise.

\section{Discussion and Conclusions}\label{sec:discussion}

The aim of this work is to check whether the test originally proposed by \cite{Gaskell88} and further developed by \cite{Dotti} is able to correctly recognize a single MBH in the presence of a non-axisymmetric BLR.

To do this, we modelled the continuum impinging on the BLR through analytical functions, either a step function or a sinusoid (Eq.~\ref{eq:step_luminosity} and \ref{eq:sin_luminosity}, respectively). We generated a disc-like BLR characterized by a spiral arm perturbation on an otherwise radial emissivity, whose profile ensures that the characteristic radius corresponds to the expectations from the observed luminosity-radius relation. The BEL profile has been shaped in wavelengths by the leading order Doppler shift associated with the local bulk velocity of each BLR element, and a local broadening associated to unresolved turbulence.
Although we did not include any relativistic effects, we compared our model with a similar implementation which also takes into account gravitational redshift, light bending, and any effect associated to radiative transfer processes, As we show in Appendix~\ref{app:model_comparison}, the most significant impact on the line profile is a shift of the spectrum toward higher wavelengths, with the blue peak increasing in height and the red peak decreasing. Notably, these differences do not compromise the results of our test, validating the robustness and reliability of FUVT.

We validated the model by testing whether it is able to approximately reproduce the luminosity-radius relation described by Eq.~\ref{eq:rad-lum relation}. We stress that ours is, to date, the first implementation encoding directly into the model the BEL response to a time dependent continuum.

After testing the physical consistency of our model, we checked the robustness of the method proposed by \cite{Dotti}. We thoroughly explored the parameter space of our model by varying the parameters of the spiral arm and computing the corresponding cross-correlation between the red part and the blue part of the resulting line. Our results rule out the possibility that spiral-like asymmetric BLRs can be mistaken by the FUVT as a MBHBs. All the different scenarios used to check the robustness of the test were characterized by a value for the cross-correlation between the red part and the blue part of the line on the order of unity, which corresponds to a correct identification of the presence of as a single MBH. 
The robustness of FUVT can be further tested, for example by modifying the BLR model, either changing the BLR dynamics (e.g. considering an eccentric disc as in \citealt{Eracleous95}) or including additional relativistic effects shaping the fine details of the BEL profile as in \cite{Chen_Halpern_Filippenko1989} and \cite{Chen_Halpern1989} (see also Appendix \ref{app:model_comparison}). However, the results obtained from this study are promising with respect to MBHB identification in EM.
 
\begin{acknowledgements} 
ES acknowledges Kaggle for providing access to their platform computational resources, including GPU support.
MD acknowledges support from ICSC – Centro Nazionale di Ricerca in High Performance Computing, Big Data and Quantum Computing, funded by European Union – NextGenerationEU, and  the Italian Ministry for Research and University (MUR) under Grant ’Progetto Dipartimenti di Eccellenza 2023-2027’ (BiCoQ).
FR acknowledges the support from the Next Generation EU funds within the National Recovery and Resilience Plan (PNRR), Mission 4 - Education and Research, Component 2 - From Research to Business (M4C2), Investment Line 3.1 - Strengthening and creation of Research Infrastructures, Project IR0000012 – “CTA+ - Cherenkov Telescope Array Plus.
M.E. acknowledges support from the U.S. National Science Foundation under award number 2205720.
M.O. acknowledges support from the U.S. National Science Foundation under award number 2205720, and the Pennsylvania State Science Achievement Graduate Fellowship Program.
MB acknowledges support provided by MUR under grant ``PNRR - Missione 4 Istruzione e Ricerca - Componente 2 Dalla Ricerca all'Impresa - Investimento 1.2 Finanziamento di progetti presentati da giovani ricercatori ID:SOE\_0163'' and by University of Milano-Bicocca under grant ``2022-NAZ-0482/B''.
\end{acknowledgements}

%
%

\bibliographystyle{aa} 
\bibliography{paper.bib}

\begin{thebibliography}{44}
\expandafter\ifx\csname natexlab\endcsname\relax\def\natexlab#1{#1}\fi

\bibitem[{{Agazie} {et~al.}(2023){Agazie}, {Anumarlapudi}, {Archibald}, {Arzoumanian}, {Baker}, {B{\'e}csy}, {Blecha}, {Brazier}, {Brook}, {Burke-Spolaor}, {Burnette}, {Case}, {Charisi}, {Chatterjee}, {Chatziioannou}, {Cheeseboro}, {Chen}, {Cohen}, {Cordes}, {Cornish}, {Crawford}, {Cromartie}, {Crowter}, {Cutler}, {Decesar}, {Degan}, {Demorest}, {Deng}, {Dolch}, {Drachler}, {Ellis}, {Ferrara}, {Fiore}, {Fonseca}, {Freedman}, {Garver-Daniels}, {Gentile}, {Gersbach}, {Glaser}, {Good}, {G{\"u}ltekin}, {Hazboun}, {Hourihane}, {Islo}, {Jennings}, {Johnson}, {Jones}, {Kaiser}, {Kaplan}, {Kelley}, {Kerr}, {Key}, {Klein}, {Laal}, {Lam}, {Lamb}, {Lazio}, {Lewandowska}, {Littenberg}, {Liu}, {Lommen}, {Lorimer}, {Luo}, {Lynch}, {Ma}, {Madison}, {Mattson}, {McEwen}, {McKee}, {McLaughlin}, {McMann}, {Meyers}, {Meyers}, {Mingarelli}, {Mitridate}, {Natarajan}, {Ng}, {Nice}, {Ocker}, {Olum}, {Pennucci}, {Perera}, {Petrov}, {Pol}, {Radovan}, {Ransom}, {Ray}, {Romano}, {Sardesai}, {Schmiedekamp}, {Schmiedekamp}, {Schmitz},
  {Schult}, {Shapiro-Albert}, {Siemens}, {Simon}, {Siwek}, {Stairs}, {Stinebring}, {Stovall}, {Sun}, {Susobhanan}, {Swiggum}, {Taylor}, {Taylor}, {Turner}, {Unal}, {Vallisneri}, {van Haasteren}, {Vigeland}, {Wahl}, {Wang}, {Witt}, {Young}, \& {Nanograv Collaboration}}]{2023ApJ...951L...8A}
{Agazie}, G., {Anumarlapudi}, A., {Archibald}, A.~M., {et~al.} 2023, \apjl, 951, L8

\bibitem[{{Amaro-Seoane} {et~al.}(2023){Amaro-Seoane}, {Andrews}, {Arca Sedda}, {Askar}, {Baghi}, {Balasov}, {Bartos}, {Bavera}, {Bellovary}, {Berry}, {Berti}, {Bianchi}, {Blecha}, {Blondin}, {Bogdanovi{\'c}}, {Boissier}, {Bonetti}, {Bonoli}, {Bortolas}, {Breivik}, {Capelo}, {Caramete}, {Cattorini}, {Charisi}, {Chaty}, {Chen}, {Chru{\'s}li{\'n}ska}, {Chua}, {Church}, {Colpi}, {D'Orazio}, {Danielski}, {Davies}, {Dayal}, {De Rosa}, {Derdzinski}, {Destounis}, {Dotti}, {Dutan}, {Dvorkin}, {Fabj}, {Foglizzo}, {Ford}, {Fouvry}, {Franchini}, {Fragos}, {Fryer}, {Gaspari}, {Gerosa}, {Graziani}, {Groot}, {Habouzit}, {Haggard}, {Haiman}, {Han}, {Istrate}, {Johansson}, {Khan}, {Kimpson}, {Kokkotas}, {Kong}, {Korol}, {Kremer}, {Kupfer}, {Lamberts}, {Larson}, {Lau}, {Liu}, {Lloyd-Ronning}, {Lodato}, {Lupi}, {Ma}, {Maccarone}, {Mandel}, {Mangiagli}, {Mapelli}, {Mathis}, {Mayer}, {McGee}, {McKernan}, {Miller}, {Mota}, {Mumpower}, {Nasim}, {Nelemans}, {Noble}, {Pacucci}, {Panessa}, {Paschalidis}, {Pfister}, {Porquet},
  {Quenby}, {Ricarte}, {R{\"o}pke}, {Regan}, {Rosswog}, {Ruiter}, {Ruiz}, {Runnoe}, {Schneider}, {Schnittman}, {Secunda}, {Sesana}, {Seto}, {Shao}, {Shapiro}, {Sopuerta}, {Stone}, {Suvorov}, {Tamanini}, {Tamfal}, {Tauris}, {Temmink}, {Tomsick}, {Toonen}, {Torres-Orjuela}, {Toscani}, {Tsokaros}, {Unal}, {V{\'a}zquez-Aceves}, {Valiante}, {van Putten}, {van Roestel}, {Vignali}, {Volonteri}, {Wu}, {Younsi}, {Yu}, {Zane}, {Zwick}, {Antonini}, {Baibhav}, {Barausse}, {Bonilla Rivera}, {Branchesi}, {Branduardi-Raymont}, {Burdge}, {Chakraborty}, {Cuadra}, {Dage}, {Davis}, {de Mink}, {Decarli}, {Doneva}, {Escoffier}, {Gandhi}, {Haardt}, {Lousto}, {Nissanke}, {Nordhaus}, {O'Shaughnessy}, {Portegies Zwart}, {Pound}, {Schussler}, {Sergijenko}, {Spallicci}, {Vernieri}, \& {Vigna-G{\'o}mez}}]{lisa2}
{Amaro-Seoane}, P., {Andrews}, J., {Arca Sedda}, M., {et~al.} 2023, Living Reviews in Relativity, 26, 2

\bibitem[{{Amaro-Seoane} {et~al.}(2017){Amaro-Seoane}, {Audley}, {Babak}, {Baker}, {Barausse}, {Bender}, {Berti}, {Binetruy}, {Born}, {Bortoluzzi}, {Camp}, {Caprini}, {Cardoso}, {Colpi}, {Conklin}, {Cornish}, {Cutler}, {Danzmann}, {Dolesi}, {Ferraioli}, {Ferroni}, {Fitzsimons}, {Gair}, {Gesa Bote}, {Giardini}, {Gibert}, {Grimani}, {Halloin}, {Heinzel}, {Hertog}, {Hewitson}, {Holley-Bockelmann}, {Hollington}, {Hueller}, {Inchauspe}, {Jetzer}, {Karnesis}, {Killow}, {Klein}, {Klipstein}, {Korsakova}, {Larson}, {Livas}, {Lloro}, {Man}, {Mance}, {Martino}, {Mateos}, {McKenzie}, {McWilliams}, {Miller}, {Mueller}, {Nardini}, {Nelemans}, {Nofrarias}, {Petiteau}, {Pivato}, {Plagnol}, {Porter}, {Reiche}, {Robertson}, {Robertson}, {Rossi}, {Russano}, {Schutz}, {Sesana}, {Shoemaker}, {Slutsky}, {Sopuerta}, {Sumner}, {Tamanini}, {Thorpe}, {Troebs}, {Vallisneri}, {Vecchio}, {Vetrugno}, {Vitale}, {Volonteri}, {Wanner}, {Ward}, {Wass}, {Weber}, {Ziemer}, \& {Zweifel}}]{lisa1}
{Amaro-Seoane}, P., {Audley}, H., {Babak}, S., {et~al.} 2017, arXiv e-prints, arXiv:1702.00786

\bibitem[{{Begelman} {et~al.}(1980){Begelman}, {Blandford}, \& {Rees}}]{1980Natur.287..307B}
{Begelman}, M.~C., {Blandford}, R.~D., \& {Rees}, M.~J. 1980, \nat, 287, 307

\bibitem[{{Bentz} {et~al.}(2009){Bentz}, {Peterson}, {Netzer}, {Pogge}, \& {Vestergaard}}]{bentz09b}
{Bentz}, M.~C., {Peterson}, B.~M., {Netzer}, H., {Pogge}, R.~W., \& {Vestergaard}, M. 2009, \apj, 697, 160

\bibitem[{{Blandford} \& {McKee}(1982)}]{Blandford82}
{Blandford}, R.~D. \& {McKee}, C.~F. 1982, \apj, 255, 419

\bibitem[{{Bogdanovi{\'c}} {et~al.}(2022){Bogdanovi{\'c}}, {Miller}, \& {Blecha}}]{Bogdanovic2022}
{Bogdanovi{\'c}}, T., {Miller}, M.~C., \& {Blecha}, L. 2022, Living Reviews in Relativity, 25, 3

\bibitem[{{Chen} \& {Halpern}(1989)}]{Chen_Halpern1989}
{Chen}, K. \& {Halpern}, J.~P. 1989, \apj, 344, 115

\bibitem[{{Chen} {et~al.}(1989){Chen}, {Halpern}, \& {Filippenko}}]{Chen_Halpern_Filippenko1989}
{Chen}, K., {Halpern}, J.~P., \& {Filippenko}, A.~V. 1989, \apj, 339, 742

\bibitem[{{De Rosa} {et~al.}(2019){De Rosa}, {Vignali}, {Bogdanovi{\'c}}, {Capelo}, {Charisi}, {Dotti}, {Husemann}, {Lusso}, {Mayer}, {Paragi}, {Runnoe}, {Sesana}, {Steinborn}, {Bianchi}, {Colpi}, {del Valle}, {Frey}, {Gab{\'a}nyi}, {Giustini}, {Guainazzi}, {Haiman}, {Herrera Ruiz}, {Herrero-Illana}, {Iwasawa}, {Komossa}, {Lena}, {Loiseau}, {Perez-Torres}, {Piconcelli}, \& {Volonteri}}]{2019NewAR..8601525D}
{De Rosa}, A., {Vignali}, C., {Bogdanovi{\'c}}, T., {et~al.} 2019, \nar, 86, 101525

\bibitem[{{Decarli} {et~al.}(2013){Decarli}, {Dotti}, {Fumagalli}, {Tsalmantza}, {Montuori}, {Lusso}, {Hogg}, \& {Prochaska}}]{Decarli13}
{Decarli}, R., {Dotti}, M., {Fumagalli}, M., {et~al.} 2013, \mnras, 433, 1492

\bibitem[{{Donnan} {et~al.}(2021){Donnan}, {Horne}, \& {Hern{\'a}ndez Santisteban}}]{Donnan2021}
{Donnan}, F.~R., {Horne}, K., \& {Hern{\'a}ndez Santisteban}, J.~V. 2021, \mnras, 508, 5449

\bibitem[{{Dotti} {et~al.}(2022){Dotti}, {Bonetti}, {D'Orazio}, {Haiman}, \& {Ho}}]{2022MNRAS.509..212D}
{Dotti}, M., {Bonetti}, M., {D'Orazio}, D.~J., {Haiman}, Z., \& {Ho}, L.~C. 2022, \mnras, 509, 212

\bibitem[{{Dotti} {et~al.}(2023{\natexlab{a}}){Dotti}, {Bonetti}, {Rigamonti}, {Bortolas}, {Fossati}, {Decarli}, {Covino}, {Lupi}, {Franchini}, {Sesana}, \& {Calderone}}]{dotti23}
{Dotti}, M., {Bonetti}, M., {Rigamonti}, F., {et~al.} 2023{\natexlab{a}}, \mnras, 518, 4172

\bibitem[{{Dotti} {et~al.}(2023{\natexlab{b}}){Dotti}, {Rigamonti}, {Rinaldi}, {Del Pozzo}, {Decarli}, \& {Buscicchio}}]{Dotti}
{Dotti}, M., {Rigamonti}, F., {Rinaldi}, S., {et~al.} 2023{\natexlab{b}}, \aap, 680, A69

\bibitem[{{Dotti} {et~al.}(2012){Dotti}, {Sesana}, \& {Decarli}}]{2012AdAst2012E...3D}
{Dotti}, M., {Sesana}, A., \& {Decarli}, R. 2012, Advances in Astronomy, 2012, 940568

\bibitem[{{EPTA Collaboration} {et~al.}(2023){EPTA Collaboration}, {InPTA Collaboration}, {Antoniadis}, {Arumugam}, {Arumugam}, {Babak}, {Bagchi}, {Bak Nielsen}, {Bassa}, {Bathula}, {Berthereau}, {Bonetti}, {Bortolas}, {Brook}, {Burgay}, {Caballero}, {Chalumeau}, {Champion}, {Chanlaridis}, {Chen}, {Cognard}, {Dandapat}, {Deb}, {Desai}, {Desvignes}, {Dhanda-Batra}, {Dwivedi}, {Falxa}, {Ferdman}, {Franchini}, {Gair}, {Goncharov}, {Gopakumar}, {Graikou}, {Grie{\ss}meier}, {Guillemot}, {Guo}, {Gupta}, {Hisano}, {Hu}, {Iraci}, {Izquierdo-Villalba}, {Jang}, {Jawor}, {Janssen}, {Jessner}, {Joshi}, {Kareem}, {Karuppusamy}, {Keane}, {Keith}, {Kharbanda}, {Kikunaga}, {Kolhe}, {Kramer}, {Krishnakumar}, {Lackeos}, {Lee}, {Liu}, {Liu}, {Lyne}, {McKee}, {Maan}, {Main}, {Mickaliger}, {Ni{\c{t}}u}, {Nobleson}, {Paladi}, {Parthasarathy}, {Perera}, {Perrodin}, {Petiteau}, {Porayko}, {Possenti}, {Prabu}, {Quelquejay Leclere}, {Rana}, {Samajdar}, {Sanidas}, {Sesana}, {Shaifullah}, {Singha}, {Speri}, {Spiewak}, {Srivastava},
  {Stappers}, {Surnis}, {Susarla}, {Susobhanan}, {Takahashi}, {Tarafdar}, {Theureau}, {Tiburzi}, {van der Wateren}, {Vecchio}, {Venkatraman Krishnan}, {Verbiest}, {Wang}, {Wang}, \& {Wu}}]{2023A&A...678A..50E}
{EPTA Collaboration}, {InPTA Collaboration}, {Antoniadis}, J., {et~al.} 2023, \aap, 678, A50

\bibitem[{{Eracleous} {et~al.}(2012){Eracleous}, {Boroson}, {Halpern}, \& {Liu}}]{Eracleous12}
{Eracleous}, M., {Boroson}, T.~A., {Halpern}, J.~P., \& {Liu}, J. 2012, \apjs, 201, 23

\bibitem[{{Eracleous} {et~al.}(1995){Eracleous}, {Halpern}, \& {Storchi-Bergmann}}]{Eracleous95}
{Eracleous}, M., {Halpern}, L., \& {Storchi-Bergmann}, T. 1995, The Astrophysical Journal, 438, 610

\bibitem[{{Gaskell}(1988)}]{Gaskell88}
{Gaskell}, C.~M. 1988, in Active Galactic Nuclei, ed. H.~R. {Miller} \& P.~J. {Wiita}, Vol. 307, 61

\bibitem[{{Gilbert} {et~al.}(1999){Gilbert}, {Eracleous}, {Filippenko}, \& {Halpern}}]{Gilbert1999}
{Gilbert}, A.~M., {Eracleous}, M., {Filippenko}, A.~V., \& {Halpern}, J.~P. 1999, in Astronomical Society of the Pacific Conference Series, Vol. 175, Structure and Kinematics of Quasar Broad Line Regions, ed. C.~M. {Gaskell}, W.~N. {Brandt}, M.~{Dietrich}, D.~{Dultzin-Hacyan}, \& M.~{Eracleous}, 189

\bibitem[{{Kaspi} {et~al.}(2000){Kaspi}, {Smith}, {Netzer}, {Maoz}, {Jannuzi}, \& {Giveon}}]{kaspi00}
{Kaspi}, S., {Smith}, P.~S., {Netzer}, H., {et~al.} 2000, \apj, 533, 631

\bibitem[{{Krolik}(2001)}]{krolik01}
{Krolik}, J.~H. 2001, \apj, 551, 72

\bibitem[{{Liu} {et~al.}(2016){Liu}, {Gezari}, {Burgett}, {Chambers}, {Draper}, {Hodapp}, {Huber}, {Kudritzki}, {Magnier}, {Metcalfe}, {Tonry}, {Wainscoat}, \& {Waters}}]{Liu2016}
{Liu}, T., {Gezari}, S., {Burgett}, W., {et~al.} 2016, \apj, 833, 6

\bibitem[{{Pancoast} {et~al.}(2014){Pancoast}, {Brewer}, {Treu}, {Park}, {Barth}, {Bentz}, \& {Woo}}]{pancoast14}
{Pancoast}, A., {Brewer}, B.~J., {Treu}, T., {et~al.} 2014, \mnras, 445, 3073

\bibitem[{{Peterson}(1993)}]{peterson93}
{Peterson}, B.~M. 1993, \pasp, 105, 247

\bibitem[{{Peterson} {et~al.}(2004){Peterson}, {Ferrarese}, {Gilbert}, {Kaspi}, {Malkan}, {Maoz}, {Merritt}, {Netzer}, {Onken}, {Pogge}, {Vestergaard}, \& {Wandel}}]{peterson04}
{Peterson}, B.~M., {Ferrarese}, L., {Gilbert}, K.~M., {et~al.} 2004, \apj, 613, 682

\bibitem[{{Popovi{\'c}}(2012)}]{Popovic2012}
{Popovi{\'c}}, L.~{\v{C}}. 2012, \nar, 56, 74

\bibitem[{{Pozo Nu{\~n}ez} {et~al.}(2023){Pozo Nu{\~n}ez}, {Gianniotis}, \& {Polsterer}}]{Nunez2023}
{Pozo Nu{\~n}ez}, F., {Gianniotis}, N., \& {Polsterer}, K.~L. 2023, \aap, 674, A83

\bibitem[{{Raimundo} {et~al.}(2020){Raimundo}, {Vestergaard}, {Goad}, {Grier}, {Williams}, {Peterson}, \& {Treu}}]{Raimundo2020}
{Raimundo}, S.~I., {Vestergaard}, M., {Goad}, M.~R., {et~al.} 2020, \mnras, 493, 1227

\bibitem[{{Reardon} {et~al.}(2023){Reardon}, {Zic}, {Shannon}, {Hobbs}, {Bailes}, {Di Marco}, {Kapur}, {Rogers}, {Thrane}, {Askew}, {Bhat}, {Cameron}, {Cury{\l}o}, {Coles}, {Dai}, {Goncharov}, {Kerr}, {Kulkarni}, {Levin}, {Lower}, {Manchester}, {Mandow}, {Miles}, {Nathan}, {Os{\l}owski}, {Russell}, {Spiewak}, {Zhang}, \& {Zhu}}]{2023ApJ...951L...6R}
{Reardon}, D.~J., {Zic}, A., {Shannon}, R.~M., {et~al.} 2023, \apjl, 951, L6

\bibitem[{{Rigamonti} {et~al.}(2025){Rigamonti}, {Severgnini, P.}, {Sottocorno, E.}, {Dotti, M.}, {Covino, S.}, {Landoni, M.}, {Bertassi, L.}, {Braito, V.}, {Cicone, C.}, {Cupani, G.}, {De Rosa, A.}, {Della Ceca, R.}, {Ighina, L.}, {Singh, J.}, \& {Vignali, C.}}]{Rigamonti2025}
{Rigamonti}, F., {Severgnini, P.}, {Sottocorno, E.}, {et~al.} 2025, \aap, 693, A117

\bibitem[{{Rodriguez} {et~al.}(2009){Rodriguez}, {Taylor}, {Zavala}, {Pihlstr{\"o}m}, \& {Peck}}]{Rodriguez09}
{Rodriguez}, C., {Taylor}, G.~B., {Zavala}, R.~T., {Pihlstr{\"o}m}, Y.~M., \& {Peck}, A.~B. 2009, \apj, 697, 37

\bibitem[{{Runnoe} {et~al.}(2015){Runnoe}, {Eracleous}, {Mathes}, {Pennell}, {Boroson}, {Sigur{\dh}sson}, {Bogdanovi{\'c}}, {Halpern}, \& {Liu}}]{Runnoe15}
{Runnoe}, J.~C., {Eracleous}, M., {Mathes}, G., {et~al.} 2015, \apjs, 221, 7

\bibitem[{{Runnoe} {et~al.}(2017){Runnoe}, {Eracleous}, {Pennell}, {Mathes}, {Boroson}, {Sigur{\dh}sson}, {Bogdanovi{\'c}}, {Halpern}, {Liu}, \& {Brown}}]{Runnoe17}
{Runnoe}, J.~C., {Eracleous}, M., {Pennell}, A., {et~al.} 2017, \mnras, 468, 1683

\bibitem[{{Storchi-Bergmann} {et~al.}(2003){Storchi-Bergmann}, {Nemmen da Silva}, {Eracleous}, {Halpern}, {Wilson}, {Filippenko}, {Ruiz}, {Smith}, \& {Nagar}}]{storchi03}
{Storchi-Bergmann}, T., {Nemmen da Silva}, R., {Eracleous}, M., {et~al.} 2003, \apj, 598, 956

\bibitem[{{Storchi-Bergmann} {et~al.}(2017){Storchi-Bergmann}, {Schimoia}, {Peterson}, {Elvis}, {Denney}, {Eracleous}, \& {Nemmen}}]{Storchi17}
{Storchi-Bergmann}, T., {Schimoia}, J.~S., {Peterson}, B.~M., {et~al.} 2017, \apj, 835, 236

\bibitem[{{Sun} {et~al.}(2018){Sun}, {Grier}, \& {Peterson}}]{Sun}
{Sun}, M., {Grier}, C.~J., \& {Peterson}, B.~M. 2018, {PyCCF: Python Cross Correlation Function for reverberation mapping studies}, Astrophysics Source Code Library, record ascl:1805.032

\bibitem[{{Tsalmantza} {et~al.}(2011){Tsalmantza}, {Decarli}, {Dotti}, \& {Hogg}}]{Tsalmantza11}
{Tsalmantza}, P., {Decarli}, R., {Dotti}, M., \& {Hogg}, D.~W. 2011, \apj, 738, 20

\bibitem[{{Vaughan} {et~al.}(2016){Vaughan}, {Uttley}, {Markowitz}, {Huppenkothen}, {Middleton}, {Alston}, {Scargle}, \& {Farr}}]{Vaughan2014}
{Vaughan}, S., {Uttley}, P., {Markowitz}, A.~G., {et~al.} 2016, \mnras, 461, 3145

\bibitem[{{Verbiest} {et~al.}(2016){Verbiest}, {Lentati}, {Hobbs}, {van Haasteren}, {Demorest}, {Janssen}, {Wang}, {Desvignes}, {Caballero}, {Keith}, {Champion}, {Arzoumanian}, {Babak}, {Bassa}, {Bhat}, {Brazier}, {Brem}, {Burgay}, {Burke-Spolaor}, {Chamberlin}, {Chatterjee}, {Christy}, {Cognard}, {Cordes}, {Dai}, {Dolch}, {Ellis}, {Ferdman}, {Fonseca}, {Gair}, {Garver-Daniels}, {Gentile}, {Gonzalez}, {Graikou}, {Guillemot}, {Hessels}, {Jones}, {Karuppusamy}, {Kerr}, {Kramer}, {Lam}, {Lasky}, {Lassus}, {Lazarus}, {Lazio}, {Lee}, {Levin}, {Liu}, {Lynch}, {Lyne}, {Mckee}, {McLaughlin}, {McWilliams}, {Madison}, {Manchester}, {Mingarelli}, {Nice}, {Os{\l}owski}, {Palliyaguru}, {Pennucci}, {Perera}, {Perrodin}, {Possenti}, {Petiteau}, {Ransom}, {Reardon}, {Rosado}, {Sanidas}, {Sesana}, {Shaifullah}, {Shannon}, {Siemens}, {Simon}, {Smits}, {Spiewak}, {Stairs}, {Stappers}, {Stinebring}, {Stovall}, {Swiggum}, {Taylor}, {Theureau}, {Tiburzi}, {Toomey}, {Vallisneri}, {van Straten}, {Vecchio}, {Wang}, {Wen}, {You},
  {Zhu}, \& {Zhu}}]{pta}
{Verbiest}, J.~P.~W., {Lentati}, L., {Hobbs}, G., {et~al.} 2016, \mnras, 458, 1267

\bibitem[{{Ward} {et~al.}(2024){Ward}, {Gezari}, {Nugent}, {Kerr}, {Eracleous}, {Frederick}, {Hammerstein}, {Graham}, {van Velzen}, {Kasliwal}, {Laher}, {Masci}, {Purdum}, {Racine}, \& {Smith}}]{Ward24}
{Ward}, C., {Gezari}, S., {Nugent}, P., {et~al.} 2024, \apj, 961, 172

\bibitem[{{Xu} {et~al.}(2023){Xu}, {Chen}, {Guo}, {Jiang}, {Wang}, {Xu}, {Xue}, {Nicolas Caballero}, {Yuan}, {Xu}, {Wang}, {Hao}, {Luo}, {Lee}, {Han}, {Jiang}, {Shen}, {Wang}, {Wang}, {Xu}, {Wu}, {Manchester}, {Qian}, {Guan}, {Huang}, {Sun}, \& {Zhu}}]{2023RAA....23g5024X}
{Xu}, H., {Chen}, S., {Guo}, Y., {et~al.} 2023, Research in Astronomy and Astrophysics, 23, 075024

\bibitem[{{Zu} {et~al.}(2011){Zu}, {Kochanek}, \& {Peterson}}]{Zu2011}
{Zu}, Y., {Kochanek}, C.~S., \& {Peterson}, B.~M. 2011, \apj, 735, 80

\end{thebibliography}

\appendix
\section{Emissivity Parameters}\label{app:emissivity}

In this Appendix we show how the BLR brightness changes with the parameters of the spiral arm model.

\begin{figure}[h]
    \begin{minipage}{\textwidth}
        \includegraphics[width=\textwidth]{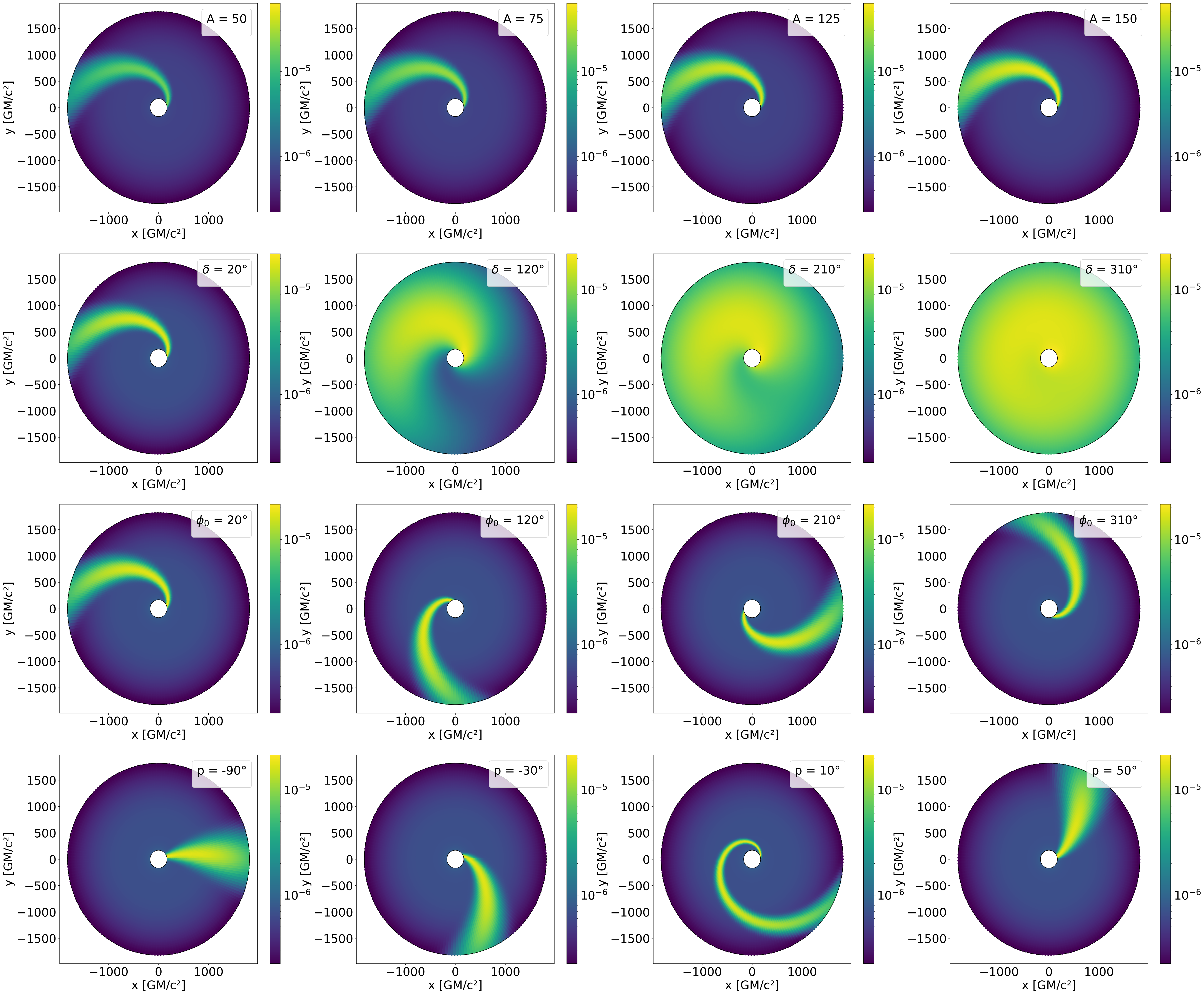}
        \caption{Effect of the emissivity parameters on the BLR emissivity pattern. Each row underline the effects of an increasing $A$, $\delta$, $\phi_0$ and $p$ respectively. The parameter in the legend is the only one varying along the row, while the others are kept constant to the following value: $A = 100$ and $\delta,\,\phi_0,\, p = 20$°.} 
        \label{fig:emissivity parameters}
    \end{minipage}
\end{figure}

\clearpage
\section{Model Comparison}\label{app:model_comparison}

\begin{figure*}[h]
    \begin{minipage}{\textwidth}
    \includegraphics[width=\textwidth]{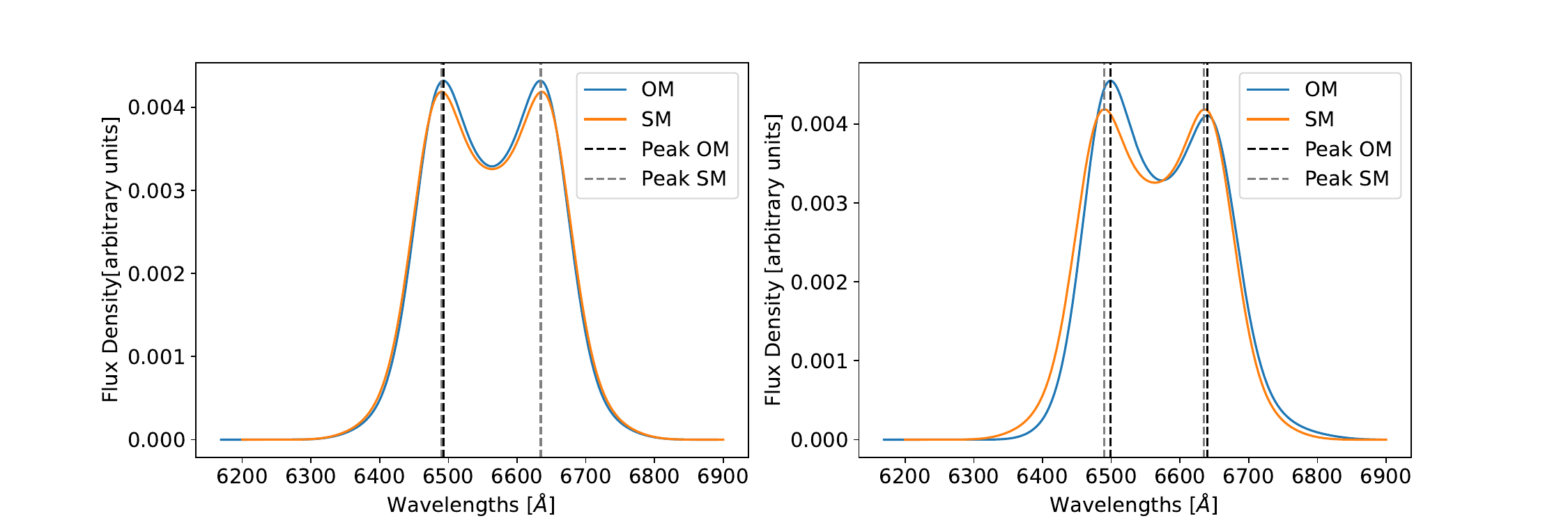}
    \caption{Comparison between the spectra obtained for the BLR without a spiral arm. On the left, both models include only the Doppler effect; on the right, OM also considers relativistic effects.}
    \label{fig:comparison_example_1}
    \end{minipage}
\end{figure*}

\begin{figure*}[h]
    \begin{minipage}{\textwidth}
    \includegraphics[width=\textwidth]{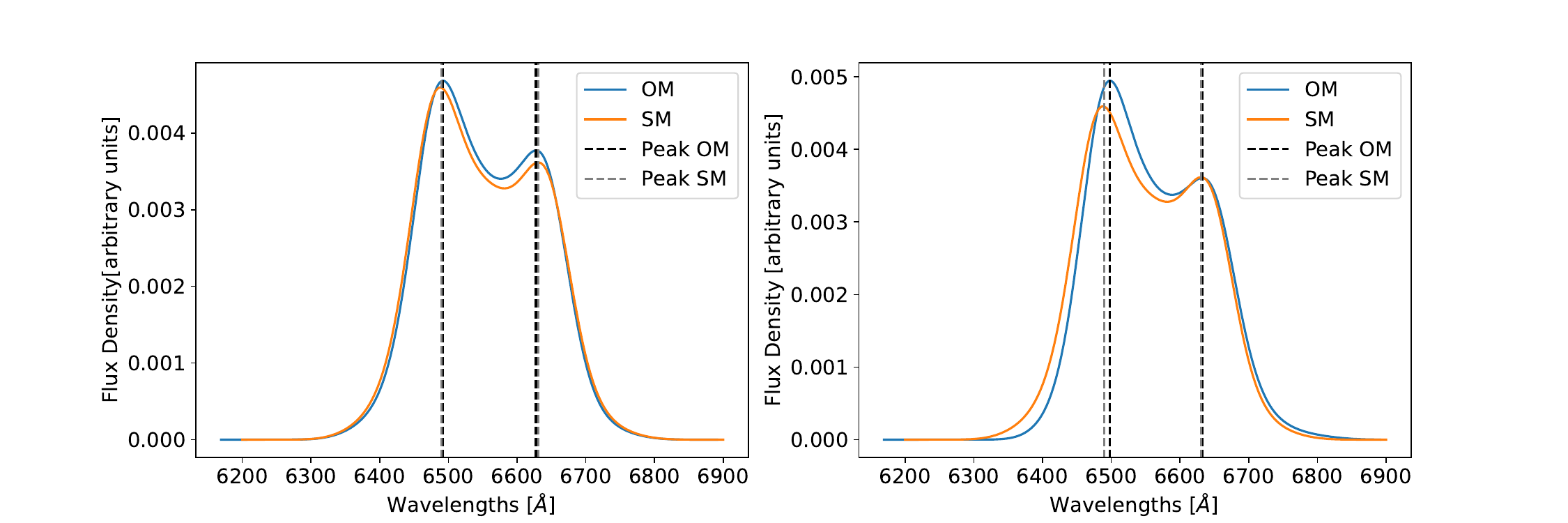}
    \caption{Comparison between the spectra obtained for the BLR with a spiral arm. On the left, both models include only the Doppler effect; on the right, OM also considers relativistic effects.}
    \label{fig:comparison_example_3}
    \end{minipage}
\end{figure*}

In this Appendix, we compare the results of our model - hereafter referred to as Sottocorno's Model (SM) - with those obtained by a model that involves an expansion at leading order of different relativistic effects. In particular, while we only consider the Doppler effect, the model (referred to as OM in the following) presented in Ogborn et al. (in prep.) also includes gravitational redshift, Doppler boosting and transverse redshift. In particular, the OM model includes the same relativistic effects as the model of \cite{Chen_Halpern_Filippenko1989} and \cite{Chen_Halpern1989}, except light bending which negligible. To illustrate the impact of relativistic effects, we consider two different scenarios: one without a spiral arm (i.e. $A = 0$) and another where the spiral arm is present (i.e. $A\neq 0$). The corresponding parameters are shown in Tab.~\ref{tab:spiral parameters example_1} and Tab.~\ref{tab:spiral parameters example_3} respectively. Both the comparisons were done by considering a constant luminosity, so that the brightness and emissivity patterns are exactly the same. 

\begin{table}[h]    
    \centering     
    \begin{minipage}[t]{0.2\textwidth}
        \centering
        \begin{tabular}{cc}
            \midrule
            Parameter & Value \\
            \midrule
            \textit{i} [deg] & 30 \\
            $\xi_{1}$ & 200 \\
            $\xi_{2}$ & 1800\\
            $\xi_{sp}$ & 200 \\
            \midrule
        \end{tabular}
    \end{minipage}
    \begin{minipage}[t]{0.2\textwidth}
        \centering
        \begin{tabular}{cc}
            \midrule
            Parameter & Value \\
            \midrule
            $\phi_0$ [deg] & - \\
            $\delta$ [deg] & - \\
            \textit{A} & 0 (no spiral) \\
            \textit{p} [deg] & - \\
            \midrule
        \end{tabular}
    \end{minipage}
    \caption{BLR geometry and emissivity parameters for the scenario without a spiral arm.}
    \label{tab:spiral parameters example_1}
\end{table}

\begin{table}[h]
    \centering    
    \begin{minipage}[t]{0.2\textwidth}
        \centering
        \begin{tabular}{cc}
            \midrule
            Parameter & Value \\
            \midrule
            \textit{i} [deg] & 30 \\
            $\xi_{1}$ & 200 \\
            $\xi_{2}$ & 1800\\
            $\xi_{sp}$ & 200 \\
            \midrule
        \end{tabular}
    \end{minipage}
    \begin{minipage}[t]{0.2\textwidth}
        \centering
        \begin{tabular}{cc}
            \midrule
            Parameter & Value \\
            \midrule
            $\phi_0$ [deg] & 120 \\
            $\delta$ [deg] & 110 \\
            \textit{A} & 2 \\
            \textit{p} [deg] & -25 \\
            \midrule
        \end{tabular}
    \end{minipage}
    \caption{BLR geometry and emissivity parameters for the scenario with a spiral arm.}
    \label{tab:spiral parameters example_3}
\end{table}

In our implementation, we divide the BLR into 175 × 175 sections and compute their contributions to the total spectrum for 175 wavelengths\footnote{This number of sections is specifically due to the GPU memory limits.}. In contrast, OM uses 1000 × 1000 sections to describe the BLR and computes the spectrum for 1000 wavelengths. Additionally, while both models use a polar coordinate system, the SM model employs a grid that is linear in both azimuth and radius, whereas the OM model applies a logarithmic scale to the radius while keeping azimuth linear. Therefore, we also compare the spectra obtained considering only the Doppler effect for both models in order to highlight the underlying differences due to the different model resolutions as well.

To give a quantitative measure of the differences, we relied on the following properties: first, we computed the first and second moments for each spectrum, given by:

\begin{equation}
    \begin{cases}
        & \langle \lambda \rangle = \frac{\sum_i \lambda_i f_i}{\sum_i f_i} \\\\
        & \mu_2 = \frac{\sum_i (\lambda_i - \langle \lambda \rangle)^2 f_i}{\sum_i f_i}
    \end{cases}
    \label{eq:moments}
\end{equation}

respectively. We then converted them into units of km/s:

\begin{equation}
    \begin{cases}
        & \langle v_{\rm{centroid}} \rangle = c\,\frac{\lambda-\lambda_e}{\lambda_e} \\
        & v_{\rm{disp}} = c\sqrt{\mu_2}/\lambda_e
    \end{cases}
    \label{eq:centroid-std dev}
\end{equation}
where we recall that $\lambda_e = 6563$ \AA. Then we compared the relative heights of the two peaks and their corresponding distances, which we defined respectively as:
\begin{equation}
    \begin{cases}
        & \mathrm{Relative \ Height} =\frac{ |f_{\rm{peak},1}-f_{\rm{peak},2}|}{\rm{max}\left(f_{\rm{peak},1}, f_{\rm{peak},2}\right)} \\
        & \mathrm{Peaks \ Distance} = c\frac{\left(\lambda_{\rm{peak},2} - \lambda_{\rm{peak},1}\right)}{\lambda_e}
    \end{cases}
    \label{eq:rel-height-peak-distance}
\end{equation}

In order to facilitate the comparison, we normalized the spectra so that the underlying area for each spectrum is equal to unity in both scenarios. 

Firstly, we show in Fig.~\ref{fig:comparison_example_1} the spectra obtained for the scenario without a spiral arm and in Tab.~\ref{tab:dopvsdop_example1} and Tab.~\ref{tab:dopvstot_example1} the resulting properties.

\begin{table}[h]
    \centering
    \begin{tabular}{lcc}
        \midrule
        \multicolumn{1}{c}{Property} & SM & OM \\
        \midrule
        $\langle v_{\rm{centroid}} \rangle$ [km/s] & 38 & 46\\
        $ \langle v_{\rm{disp}} \rangle $ [km/s]& 3814 & 3726\\
        Relative Height & 6.0$\cdot 10^{-4}$ & 6.5$\cdot 10^{-5}$ \\
        Peaks Distance [km/s] & 6620 & 6441 \\
        \bottomrule
    \end{tabular}
    \caption{Properties comparison between the two models for example given by Tab.~\ref{tab:spiral parameters example_1}, considering only Doppler effect in OM.}
    \label{tab:dopvsdop_example1}
\end{table}

\begin{table}[h]
    \centering
    \begin{tabular}{lcc}
        \midrule
        \multicolumn{1}{c}{Property} & SM & OM \\
        \midrule
        $ \langle v_{\rm{centroid}} \rangle$ [km/s] & 38 & 402\\
        $ \langle v_{\rm{disp}} \rangle $ [km/s] & 3814 & 3767\\
        Relative Height & 6.0$\cdot 10^{-4}$ & 9.7$\cdot 10^{-2}$\\
        Peaks Distance  [km/s] & 6620 & 6441\\
        \bottomrule
    \end{tabular}
    \caption{Properties comparison between the two models, including relativistic effects in OM, for the example in Tab.~\ref{tab:spiral parameters example_1}.}
    \label{tab:dopvstot_example1}
\end{table}

As we can see, while the scenario with only the Doppler effect included for both models remains comparable, the inclusion of relativistic effects causes a shift toward higher wavelengths and alters the relative height of the two peaks. Specifically, the blue peak becomes more pronounced, while the red peak diminishes. This results in a higher relative height compared to the scenario considering only Doppler effects. This also leads to a shift in the computed centroid, while the standard deviation and the peaks distance are barely affected.

Then, we show in Fig.~\ref{fig:comparison_example_3} the spectra obtained for the scenario with a spiral arm and in Tab.~\ref{tab:dopvsdop_example3} and Tab.~\ref{tab:dopvstot_example3} the resulting properties.

\begin{table}[h]
    \centering
    \begin{tabular}{lcc}
        \midrule
        \multicolumn{1}{c}{Property} & SM & OM \\
        \midrule
        $ \langle v_{\rm{centroid}} \rangle$ [km/s] & -484  & -431 \\
        $ \langle v_{\rm{disp}} \rangle $ [km/s]& 3787 & 3682\\
        Relative Height & 0.21 & 0.19\\
        Peaks Distance  [km/s] & 6436 & 6177\\
        \bottomrule
    \end{tabular}
    \caption{Properties comparison between the two models for example given by Tab.~\ref{tab:spiral parameters example_3}, considering only Doppler effect in OM.}
    \label{tab:dopvsdop_example3}
\end{table}

As we can see, also in this case the inclusion of relativistic effects causes a shift toward higher wavelengths and modifies the relative height of the two peaks, while the two models remain comparable if considering only Doppler. Since the impact of relativistic effects on the spectrum is comparable to the one described in the previous cases, this also results in a higher relative height compared to the scenario considering only Doppler effects and a shift on the computed centroid. However, in this case, we also observe that the differences in both the standard deviation and the peak distances diminish. 

\begin{table}[h]
    \centering
    \begin{tabular}{lcc}
        \midrule
        \multicolumn{1}{c}{Property} & SM & OM \\
        \midrule
        $ \langle v_{\rm{centroid}} \rangle$ [km/s] & -484 & -82 \\
        $ \langle v_{\rm{disp}} \rangle $ [km/s]& 3787 & 3695\\
        Relative Height & 0.21 & 0.27\\
        Peaks Distance  [km/s] & 6436 & 6144\\
        \bottomrule
    \end{tabular}
    \caption{Properties comparison between the two models, including relativistic effects in OM, for the example in Tab.~\ref{tab:spiral parameters example_3}.}
    \label{tab:dopvstot_example3}
\end{table}

Our tests confirm that including relativistic effects does not affect our results or change the conclusions of the FUVT test on the data. The key findings remain consistent in both cases, showing that, while relativistic corrections exist, they are only a second-order effect and do not significantly influence the overall outcome of the test.

\end{document}